\documentclass[prc,a4paper,onecolumn,showpacs,showkeys,superscriptaddress]{revtex4}
\usepackage[english]{babel}
\usepackage{color,graphicx,rotating}
\usepackage[left=1cm,top=2cm,right=1cm,bottom=2cm,nohead,nofoot]{geometry}
\usepackage{longtable}
\usepackage{caption}

\newcommand{\ft}{\mathcal{F}t}
\newcommand{\br}{B\!R}
\newcommand{\BR}{B\!R}

\usepackage{subfigure}
\usepackage{hyperref}

\usepackage{epsfig}
\usepackage{graphicx}
\usepackage{graphics}
\usepackage{amsmath}

\begin{document}


\title{$\mathcal{F} t$ values of the $T = 1/2$ mirror $\beta$ transitions}

\author{N. Severijns}%
\email{nathal.severijns@fys.kuleuven.be}%
\author{M. Tandecki}%
\author{T. Phalet}%
\affiliation{K.U.Leuven, Instituut voor Kern- en Stralingsfysica,
Celestijnenlaan 200D, B-3001 Leuven, Belgium}%
\author{I.S. Towner}%
\affiliation{Cyclotron Institute, Texas A \& M University, College Station,
Texas 77843, USA}%


\begin{abstract}
A complete survey is presented of all half-life and
branching-ratio measurements related to the isospin $ T = 1/2 $
mirror $\beta$ transitions ranging from $^3$He to $^{83}$Mo. No
measurements are ignored, although some are rejected for cause.
Using the decay energies obtained in the 2003 Mass Evaluation
experimental $ ft $ values are then determined for the transitions
up to $^{45}$V. For the first time also all associated theoretical
corrections needed to convert these results into "corrected"
$\mathcal{F} t$ values, similar to the superallowed $0^+
\rightarrow 0^+$ pure Fermi $\beta$ transitions, were calculated.
Precisions of the resulting values are in most cases between 0.1\%
and 0.4\%. These $\mathcal{F} t^{mirror}$ values can now be used
to extract precise weak interaction information from past and
ongoing correlation measurements in the beta decay of the $T =
1/2$ mirror $\beta$ transitions.
\end{abstract}

%






\maketitle


%


\section{Introduction}

In the past, several experiments in nuclear $\beta$-decay
searching for non-Standard Model contributions to the weak
interaction were performed with $T = 1/2$ mirror nuclei
\cite{calaprice75,schreiber83,garnett88,severijns89,masson90,converse93,melconian07,scielzo03,scielzo04,iacob06,vetter08}.
Whereas originally the accuracy of these measurements was still
rather limited (at best 2 \%), first precision results were
recently obtained with $^{21}$Na
\cite{scielzo03,scielzo04,iacob06,vetter08} while several other
experiments are ongoing (with $^{35}$Ar \cite{beck03} and $^{37}$K
\cite{behr05}) or in preparation ($^{19}$Ne
\cite{berg03,broussard05} and $^{21}$Na \cite{sohani06}). In order
to extract reliable information from such measurements, precise
knowledge of the $ft$ value of the mirror transition under
investigation is required. We have therefore performed a thorough
survey of all data in the literature related to the $ft$ values of
the $T = 1/2$ mirror $\beta$ transitions and calculated the $ft$
values for the cases up to $^{45}$V, thereby updating the previous
work of Raman {\it et al.} \cite{raman78}.

On the experimental side, half-lives, $t_{1/2}$, branching ratios,
$BR$, and $Q_{EC}$ values are required for the determination of
$ft$ values. As for the first two, the literature was searched and
data were evaluated, leading to adopted values for each isotope.
The $Q_{EC}$ values were taken from the 2003 Mass Evaluation
\cite{audi03}. Since for most nuclei up to $A \approx$ 40 the
experimental data turned out to be sufficiently precise to yield
$ft$ values with a precision at the few 10$^{-3}$ level we decided
to perform, for the first time for these mirror $\beta$
transitions, a full analysis of all radiative and nuclear
structure corrections leading to the corrected $\mathcal{F} t$
values. Up to now such complete evaluation of the $\mathcal{F} t$
value was only carried out for the superallowed $0^+ \rightarrow
0^+$ pure Fermi $\beta$ transitions \cite{hardy05}. For all $T =
1/2$ mirror nuclei up to $^{45}$V $\mathcal{F} t$ values with a
precision ranging from 0.10 \% to about 2.3 \% were obtained. For
the heavier nuclei experimental data are either not available or
not sufficiently precise. Nevertheless, all experimental data
reported in the literature are listed here.

In a first section the equation for the $ft$ value of an allowed
$\beta$ transition, including all corrections, is derived. From
this the equation for the $\mathcal{F} t$ value for the $T = 1/2$
mirror $\beta$ transitions is then deduced. The next section
explains the selection and treatment of the experimental data,
while the last section deals with the $\mathcal{F} t$ values
themselves. At the end of this paper tables are given that list
all experimental data and adopted values leading to the
$\mathcal{F} t$ values of the $T = 1/2$ mirror transitions, the
values for the different correction factors applied for the nuclei
up to $^{45}$V and, finally, the derived results for the
$\mathcal{F} t^{mirror}$ values.

\section{Formalism}

The decay rate for an allowed $\beta$-decay from an unpolarized
nucleus is written \cite{jackson57}
\begin{equation}
d\Gamma = d\Gamma_0 ~ \xi ~ \biggl[ 1 + \frac{\gamma }{W} b
\biggr] , \label{eq:Gamma}
\end{equation}
\noindent with
\begin{equation}
d\Gamma_0 = \frac{G_F^2 V_{ud}^2}{ (2 \pi)^5 } ~ \frac{1}{(m_e
c^2)^5} ~ F(\pm Z, W)~ S(\pm Z,W)~ (W - W_0)^2
~p~W~dW~d\Omega_e~d\Omega_\nu , \label{eq:Gamma0}
\end{equation}
\noindent where $W$ is the total electron energy in electron
rest-mass units, $W_0$ its maximum value, $p = \sqrt{W^2-1}$ its
momentum and $m_e c^2$ the electron rest mass. Further, $\gamma =
\sqrt{1-(\alpha Z)^2}$, with $\alpha$ the fine structure constant
and $Z$ the charge of the daughter nucleus (taken positive for
electron emission, negative for positron emission), $G_F$ is the
fundamental weak interaction coupling constant taken from muon
decay, $G_F/(\hbar c)^3 =  (1.16639 \pm 0.00001) \times 10^{-5}$
GeV$^{-2}$, $V_{ud}$ is the up-down quark-mixing element of the
Cabibbo-Kobayashi-Maskawa (CKM) matrix, $F(\pm Z, W)$ the
Fermi-function, and $S(\pm Z, W)$ is the shape-correction function
the value of which is unity in the allowed approximation, but
whose value differs weakly from one when this approximation is
relaxed. In addition, we define
\begin{equation}
\xi = 2 \left[ M_F^2 C_V^2 + M_{GT}^2 C_A^2 \right] ,
\label{eq:xi}
\end{equation}
\noindent where $M_F$ and $M_{GT}$ are the Fermi and Gamow-Teller
matrix elements respectively, and $C_V$ and $C_A$ are the strength
of the weak vector and axial-vector interactions (in units of
$G_F$) as defined in the Hamiltonian of Jackson, Treiman and Wyld
\cite{jackson57}. We have assumed maximal parity violation for V-
and A-currents. Finally, $b$ is the Fierz interference term
\cite{jackson57}. The mean lifetime $\tau$ of the decaying state
is $\hbar /\Gamma$, which after integrating over neutrino and
electron directions, yields
\begin{equation}
\hbar / \tau = \int{d \Gamma} = \int  \frac{G_F^2 V_{ud}^2}{ 2
\pi^3 } ~ \frac{1}{(m_e c^2)^5} ~ \xi ~ F(\pm Z, W)~S(\pm Z,W) ~
(W - W_0)^2 ~p~W~ \biggl[ 1 + \frac{\gamma }{W} b \biggr] dW .
\label{eq:invtau}
\end{equation}
\noindent We isolate the partial half-life $t$ by correcting for
electron capture competition, $P_{EC}$, and selecting the
branching ratio, $BR$, for the particular transition under study,
to obtain
\begin{equation}
1 / t = \frac{G_F^2 V_{ud}^2}{2 K} ~ \xi ~ f ~ b^\prime ,
\label{eq:invt}
\end{equation}
\noindent with
\begin{equation}
t = \ln 2 ~ \tau \biggl( \frac{ 1 + P_{EC} } {BR} \biggr) ~ ,
\label{eq:tECBR}
\end{equation}
\noindent and
\begin{equation}
K/(\hbar c)^6 = \frac{2 \pi^3 ~\ln 2 ~ \hbar } {(m_ec^2)^5} =
(8120.278 \pm 0.004) \times 10^{-10}~{\rm GeV}^{-4}~{\rm s} .
\label{eq:K}
\end{equation}
\noindent The statistical rate function, $f$, and the Fierz
correction factor, $b^{\prime}$, are defined as
\begin{eqnarray}
f & = & \int F(\pm Z, W)~S(\pm Z,W)~( W - W_0 )^2~p~W~dW
\label{eq:f}
\\
b^\prime & = & 1 + \langle \frac{\gamma } {W} \rangle b ~ .
\label{eq:bprime}
\end{eqnarray}
where
\begin{equation}
\langle \frac{\gamma }{W} \rangle = \frac{1}{f} \int F(\pm
Z,W)~S(\pm Z,W)~ (W-W_0)^2~p~W~\frac{\gamma }{W}~dW .
\label{eq:goverW}
\end{equation}
Inserting these definitions into Eq.~(\ref{eq:invt}), we come to
our principal result
\begin{eqnarray}
ft & = & \frac{2 K}{G_F^2 ~ V_{ud}^2} \frac{1}{\xi}
\frac{1}{b^{\prime}} ,
\nonumber \\
& = & \frac{K}{G_F^2~V_{ud}^2} \frac{1}{ \left [ M_F^2 C_V^2 +
M_{GT}^2 C_A^2 \right ]} \frac{1}{b^{\prime}} . \label{eq:ft}
\end{eqnarray}

We now introduce two classes of small corrections:  those due to
radiative processes that go undetected in the experiment, and
those due to isospin not being an exact symmetry in nuclei.
Details on the nature of these corrections can e.g. be found in
ref. \cite{hardy90}. We discuss the radiative corrections first.
These are divided into terms that depend on the nucleus in
question ('outer radiative correction'), $\delta_R$, and those
that do not ('inner radiative correction'), $\Delta_R$:
\begin{equation}
1 + RC = (1 + \delta_R) (1 + \Delta_R ) . \label{eq:RadCorr}
\end{equation}
The nuclear-dependent term can be further divided into those
pieces that depend trivially on the nucleus, $\delta_R^{\prime}$
(depending only on $Z$ and $W_0$), and those that require a
detailed nuclear-structure calculation, $\delta_{NS}$:
\begin{equation}
1 + RC = (1 + \delta_R^{\prime}) (1 + \delta_{NS}) ( 1 + \Delta_R
) . \label{eq:RCNS}
\end{equation}
The $\delta_R^{\prime}$ term is mainly obtained from a standard
QED calculation that has been completed to orders $\alpha$ and $Z
\alpha^2$ and estimated to order $Z^2 \alpha^3$ \cite{sirlin86,
jaus87,sirlin87}.  These three contributions we will call
$\delta_1$, $\delta_2$ and $\delta_3$ respectively
\begin{equation}
\delta_R^{\prime} = \delta_1 + \delta_2 + \delta_3
+\delta_{\alpha^2} , \label{eq:RC123}
\end{equation}
while the $\delta_{\alpha^2}$-term is a leading log extrapolation
of a low-energy term in the evaluation of the inner radiative
correction $\Delta_R$ \cite{marciano06} that turned out to be
weakly nucleus-dependent and was therefore shifted from the inner
radiative correction to the outer one \cite{towner08}. All four
contributions in Eq.~\ref{eq:RC123} are the same for both Fermi
and Gamow-Teller transitions. By contrast, the contributions
$\delta_{NS}$ and $\Delta_R$ differ between Fermi and Gamow-Teller
transitions and so their notation will include a superscript of
$V$ or $A$ as required. Details of the calculation of
$\delta_{NS}$ can be found in refs. \cite
{jaus90,barker92,towner92,towner94,towner08}. The
nucleus-independent radiative correction $\Delta_R$ was originally
evaluated by Marciano and Sirlin \cite{marciano86} and Sirlin
\cite{sirlin95}, yielding $\Delta_R$ = 2.40(8) \% and has recently
been addressed again by Marciano and Sirlin \cite{marciano06}
leading to the new value $\Delta_R$ = (2.361 $\pm$ 0.038)\%, in
agreement with the previous  value, but a factor of about two more
precise. The reduction of the central value by approximately
0.04\% is due to the fact that the aforementioned term
$\delta_{\alpha^2}$ was shifted from the inner radiative
correction to the outer one.

The Fermi matrix element in the isospin-symmetry limit is
precisely known -- it is given in terms of an isospin
Clebsch-Gordan coefficient.  In practice, however, nuclei are
impacted by Coulomb and other charge-dependent forces that weakly
break the isospin symmetry. So we write
\begin{equation}
M_F^2 = |M_F^0|^2 ( 1 - \delta_C^V ) , \label{eq:MF2}
\end{equation}
\noindent where $\delta_C^V$ is the isospin-symmetry breaking
correction in Fermi transitions \cite{ormand95,towner02} and
$|M_F^0|^2$ is the isospin symmetry limit value of the matrix
element squared given by $|M_F^0|^2 = 2$ for $T=1 \rightarrow T=1$
transitions, and $|M_F^0|^2 = 1$ for $T=1/2 \rightarrow T=1/2$
transitions.  By contrast, the Gamow-Teller matrix element is {\em
not} known in the isospin symmetry limit. Nevertheless, to
maintain a consistency in the equations, we write
\begin{equation}
M_{GT}^2 = |M_{GT}^0|^2 ( 1 - \delta_C^A ) \label{eq:MGT2}
\end{equation}
although separate values of the symmetry-limit matrix element,
$M_{GT}^0$, and the symmetry-breaking correction, $\delta_C^A$,
are not required for the development here. The isospin-symmetry
breaking correction in Fermi transitions, $\delta_C^V$, is
typically separated into two components \cite{towner08}
\begin{equation}
\delta_C^V = \delta_{C1}^V + \delta_{C2}^V , \label{eq:dc12}
\end{equation}
\noindent where the first term quantifies the impact of
charge-dependent configuration mixing leading to differing wave
functions for the parent and daughter nuclei, while the second
term accounts for the differences in the single-particle neutron
and proton radial wave functions, which cause the radial overlap
integral of the parent and daughter nucleus to be less than unity.

Including now all corrections, and noting the shape-correction
function $S(\pm Z,W)$ in the statistical rate function differs
between Fermi and Gamow-Teller transitions, we have (setting
$b^{\prime} = 1$)
\begin{equation}
t = \frac{ K}{G_F^2 ~ V_{ud}^2} ~ \frac{1}{\left( 1 +
\delta_R^{\prime} \right)  \left[ f_V |M_F^0|^2 \left( 1 +
\delta_{NS}^V - \delta_C^V \right) C_V^2 \left( 1 + \Delta_R^V
\right) + f_A |M_{GT}^0|^2 \left( 1 + \delta_{NS}^A - \delta_C^A
\right)
 C_A^2 \left( 1 + \Delta_R^A \right) \right]} .
\label{eq:ft_general}
\end{equation}
\noindent For the superallowed $0^+ \rightarrow 0^+$ pure Fermi
transitions, with $|M_F^0|^2$ = 2 and $M_{GT}^0$ = 0, one then has
\begin{equation}
f_{V} t^{0^+ \rightarrow 0^+} = \frac{K}{2 G_F^2 ~ V_{ud}^2} ~
\frac{1}{\left( 1 + \delta_R^{\prime} \right) \left( 1 +
\delta_{NS}^V - \delta_C^V \right) C_V^2 \left( 1 + \Delta_R^V
\right) } \label{eq:ft00}
\end{equation}
or
\begin{equation}
\mathcal{F} t^{0^+ \rightarrow 0^+} \equiv f_V t^{0^+ \rightarrow
0^+} \left( 1 + \delta_R^{\prime} \right) \left( 1 + \delta_{NS}^V
- \delta_C^V \right) = \frac{K}{2 G_F^2 ~ V_{ud}^2 ~ C_V^2 \left(
1 + \Delta_R^V \right) } . \label{eq:Ft00}
\end{equation}
\noindent For a mixed Fermi and Gamow-Teller transition, we can
recast Eq.~(\ref{eq:ft_general}) into the form
\begin{eqnarray}
f_V t (1 + \delta_R^{\prime}) (1+ \delta_{NS}^V - \delta_C^V ) & =
& \frac{K}{G_F^2 V_{ud}^2} \frac{1}{|M_F^0|^2 C_V^2 (1 +
\Delta_R^V) \left( 1 + \frac{f_A}{f_V} \rho^2 \right) } ,
\nonumber \\
& = & \frac{2 \mathcal{F} t^{0^+ \rightarrow 0^+}}
{|M_F^0|^2~\left( 1 + \frac{f_A}{f_V} \rho^2 \right)} ,
\label{eq:fvt}
\end{eqnarray}
where a mixing ratio is defined as
\begin{equation}
\rho = \frac{C_A M_{GT}^0}{C_V M_F^0} \left( \frac{(1 +
\delta_{NS}^A - \delta_C^A)(1 + \Delta_R^A)} {(1 + \delta_{NS}^V -
\delta_C^V)(1 + \Delta_R^V)} \right)^{1/2} \simeq \frac{C_A
M_{GT}^0}{C_V M_F^0} . \label{eq:rho}
\end{equation}
\noindent Lastly, restricting our attention to the $T = 1/2$
mirror $\beta$-transitions, for which $|M_F^0|^2$ = 1,
Eq.~(\ref{eq:fvt}) reduces to
\begin{equation}
\mathcal{F} t^{mirror} \equiv f_V t (1 + \delta_R^{\prime}) (1+
\delta_{NS}^V - \delta_C^V )  = \frac{2 \mathcal{F} t^{0^+
\rightarrow 0^+}} {\left( 1 + \frac{f_A}{f_V} \rho^2 \right)} .
\label{eq:master}
\end{equation}
This is our master equation.  Our goal now is to extract values of
the mixing ratio squared $\rho^2$ using data on the partial
half-lives, $t$, for mirror transitions in odd-mass nuclei.   To
this end we need apart from experimental data also calculations of
the statistical rate function, $f_V$ and the ratio $f_A/f_V$, the
nucleus-dependent radiative corrections, $\delta_R^{\prime}$ and
$\delta_{NS}^V$, and the isospin-symmetry breaking correction,
$\delta_C^V$. Further, we take the current best value of
$\mathcal{F}t^{0^+ \rightarrow 0^+}$ from the most recent work of
Towner and Hardy \cite{towner08}.

\section{Experimental data}

To determine the $ft$ value for a $\beta$ transition three
measured quantities are required: the half-life, $t_{1/2}$, of the
parent state, the branching ratio, $BR$, of the particular
transition of interest, and the total transition energy, $Q_{EC}$.
The half-life and the branching ratio combine to yield the partial
half-life, $t$, (Eq.~(\ref{eq:tECBR})), whereas the $Q_{EC}$ value
is required to determine the statistical rate function, $f$,
(Eq.~(\ref{eq:f})). In our treatment of the data all half-life and
branching ratio measurements published before January 2008 are
considered. Since the evaluation of the $Q_{EC}$ values from
different types of measurements would be too vast a project in
itself it was decided to rely for these on the very extended 2003
Mass Evaluation \cite{audi03}. Half-life and branching ratio data
are available for mirror nuclei up to $^{83}$Mo. All original
experimental data were checked in detail. In Tables~I and
~\ref{branchings} we present all measured values for the half-life
and the branching ratio that were used in our analysis. References
to these data are listed in Tables~\ref{usedref_half-life}
and~\ref{usedref_BR}.  Each datum appearing in these tables is
attributed to its original journal reference via an alphanumeric
code comprising the initial two letters of the first author's name
and the last two digits of the publication date. If data were
obviously wrong they were rejected. All rejected data are listed
in Tables~\ref{unusedref_half-life} and~\ref{unusedref_BR}, with
the reason for this rejection.

Similar evaluation principles and statistical procedures as those
that are adopted for the analysis of the superallowed $0^+
\rightarrow 0^+$ pure Fermi transitions \cite{hardy05} were used.
Thus, of the surviving results, only those with uncertainties that
are within a factor of 10 of the most precise measurement for each
quantity were retained for averaging in the tables.

The statistical procedures followed in analyzing the tabulated
data are based on those used by the Particle Data Group in their
periodic reviews of particle properties (e.g. Ref. \cite{yao06}).
In the tables and throughout this work, "error bars" and
"uncertainties" always refer to plus/minus one standard deviation
(68\% confidence level).

For a set of $N$ independent measurements, $x_i \pm \delta x_i$,
of a particular quantity, a Gaussian distribution is assumed, the
weighted average being calculated according to the equation

\begin{equation}
\bar x \pm \delta \bar x = \frac{{\sum\limits_i {w_i x_i }
}}{{\sum\limits_i {w_i } }} \pm \left( {\sum\limits_i {w_i } }
\right)^{ - 1/2}   , \label{eq:wave}
\end{equation}

\noindent where

\begin{equation}
w_i = 1 / (\delta x_i)^2
\end{equation}

\noindent and the sums extend over all $N$ measurements. For each
average the $\chi^2$ is also calculated and a scale factor, $S$,
determined from

\begin{equation}
S = [ \chi^2 / (N-1) ]^{1/2}. \label{scale}
\end{equation}

\noindent This factor is then used to establish the quoted
uncertainty. If $S \leq$ 1, the value of $\delta \bar x$ from
Eq.~(\ref{eq:wave}) is left unchanged. If $S >$1 and the input
$\delta x_i$ are all about the same size, then $\delta \bar x$ is
increased by the factor $S$, which is equivalent to assuming that
all the experimental errors were underestimated by the same
factor. Finally, if $S >$1 but the $\delta x_i$ are of widely
varying magnitudes, $S$ is recalculated with only those results
for which $\delta x_i \leq 3 N^{1/2} \delta \bar x $ being
retained; the recalculated scale factor is then applied in the
usual way. In all three cases, no change is made to the original
average $\bar x$ calculated with Eq.~(\ref{eq:wave}).

Adopted values for the half-life and the branching ratio are
listed in Table~\ref{adopted_t_BR}, together with the calculated
electron-capture fraction, $P_{EC}$, the deduced partial
half-life, $t$, (cf. Eq.~(\ref{eq:tECBR})) and the $Q_{EC}$ value
from ref.~\cite{audi03}. The $P_{EC}$ values were obtained from
the tables of Bambynek $\textit{et al.}$ \cite{bambynek77} and
Firestone \cite{firestone96}. No errors were assigned to these
$P_{EC}$ values as they are expected to be accurate to a few parts
in 100 \cite{hardy05,bambynek77} such that they do not contribute
perceptibly to the overall uncertainties.

\section{The $\mathcal{F} t^{mirror}$ values}

Having surveyed the experimental data we can now turn to the
determination of the $ft$ values. The statistical rate function,
$f$, for each transition was calculated using the procedure and
the code described in \cite{hardy05}. Results appear in column 2
of Table~\ref{calculated quantities}. To obtain $\mathcal{F}
t^{mirror}$ values according to Eq.~(\ref{eq:master}) we must
still deal with the small correction terms. The values for the
nucleus dependent radiative correction $\delta_R^{\prime} =
\delta_1 + \delta_2 + \delta_3 + \delta_{\alpha^2}$ are listed in
columns 5 to 9 of Table~\ref{calculated quantities}. Similar to
the superallowed Fermi $\beta$ decays we have assigned an
uncertainty equal to the $\delta_3$ term as an estimate of the
error made in stopping the calculations at the order
$Z_2\alpha^3$. Finally, one still has to deal with the
nuclear-structure dependent corrections $\delta_C^V =
\delta_{C1}^V + \delta_{C2}^V$ and $\delta_{NS}^V$. Two of these
corrections, $\delta_{NS}^V$ and $\delta_{C1}^V$, are very
sensitive to the details of the shell-model calculation used in
their evaluation. Fortunately, these two terms are also the
smallest of the corrections we need in Eq.~(\ref{eq:master}). We
have mounted shell-model calculations using standard effective
interactions and modest-size model spaces to evaluate them
following exactly the same procedures as discussed in
ref.\cite{towner02}. Further we assigned a generous error to
account for their inherent model dependence.  Less dependent on
nuclear structure is the larger radial overlap correction,
$\delta_{c2}$.  Here we are guided by the recent work of Towner
and Hardy \cite{towner08}, who pointed out the importance of
including 'core' orbitals in the shell-model evaluation of
spectroscopic amplitudes.  A decision has to be made as to which
core orbitals should be included in the active model space. Towner
and Hardy's criterion is that experimental neutron pick-up
reactions should observe strong spectroscopic factors for the
orbitals in question.  We have followed this criterion in
obtaining our values for $\delta_{c2}$.  All these corrections are
listed in columns 10 to 12 in Table~\ref{calculated quantities}
with their sum in column 13.  In total, these nuclear-structure
dependent corrections are of order one percent or less.

One other quantity that depends weakly on a shell-model
calculation is the ratio $f_A/f_V$.  Here a modest shell-model
calculation is sufficient. We can also use these shell-model
calculations to determine the relative sign of the Fermi and
Gamow-Teller matrix elements, which can then be taken as the sign
of $\rho$ in Eq.~(\ref{eq:rho}). Finally, the resulting
$\mathcal{F} t^{mirror}$ values and corresponding values for
$\rho$ (using $\mathcal{F} t^{0^+ \rightarrow 0^+}$ = (3071.4
$\pm$ 8) s \cite{towner08}) are recorded in Table~\ref{Ft and
rho}. As can be seen, for most of the nineteen transitions the
precision on the $\mathcal{F} t^{mirror}$ value is better than 1
\%, except for $^{43}$Ti and $^{45}$V, while it is even better
than 0.3 \% in nine cases. The highest precision is reached for
$^3$H, $^{13}$N and $^{35}$Ar.

In figure 1 the fractional uncertainties attributed to each
experimental and theoretical input factor that contributes to the
final $\mathcal{F} t^{mirror}$ value are shown in the form of a
histogram for all nineteen transitions. Clearly, to bring all
contributions at the level of 1 part in 1000 or better, new and
more precise measurements of the half-lives, $t_{1/2}$, are
required for almost all transitions. Better $Q_{EC}$ values are
needed for almost half of the transitions, i.e. $^{11}$C,
$^{15}$O, $^{21}$Na, $^{23}$Mg, $^{31}$S, $^{39}$Ca, $^{43}$Ti and
$^{45}$V, while more precise measurements of the branching ratio,
$BR$, are needed for $^{23}$Mg, $^{33}$Cl, $^{37}$K, $^{43}$Ti and
$^{45}$V. The theoretical corrections, $\delta_R$ and
$\delta_C-\delta_{NS}$ contribute less than 1 part in 1000 to the
final $\mathcal{F} t^{mirror}$ values in all cases except
$^{43}$Ti and $^{45}$V.

\section{Standard model values for the $\beta$ decay correlation coeficients}

With these values for $\rho$ we can now calculate the standard
model values for correlation coefficients in $\beta$ decay
\cite{jackson57} that are of interest to search for physics beyond
the standard electroweak model (e.g.
\cite{herczeg01,erler05,severijns06,abele08}). The standard model
assumes only vector and axial-vector interactions with maximal
parity violation. In addition it is expected that the effects due
to CP (or T) violation are negligible in the light quark sector at
the present level of precision. These assumptions result in the
conditions $C^\prime_V = C_V, C^\prime_A = C_A, C_S = C^\prime_S =
C_T = C^\prime_T = 0$ and $Im(C^\prime_i) = Im(C_i) = 0$ for $i =
V, A$. Neglecting Coulomb as well as induced recoil effects one
then obtains (the upper sign is for $\beta^-$ decay, the lower
sign for $\beta^+$ decay), for the $\beta$-neutrino angular
correlation coefficient

\begin{equation}
\label{eqn:acorr-SM} a_{SM} = \frac{1 - \rho^2/3}{1 + \rho^2},
\end{equation}

\noindent for the $\beta$ asymmetry parameter

\begin{eqnarray}
\label{eqn:Acoeff-SM} A_{SM}  =  \frac{ \mp \lambda_{J^\prime J}
\rho^2 - 2 \delta_{J^\prime J} \sqrt{\frac{J}{J+1}} \rho } {1 +
\rho^2},
\end{eqnarray}

\noindent for the neutrino asymmetry parameter

\begin{eqnarray}
\label{eqn:Bcoeff-SM} B_{SM}  =  \frac{ \pm \lambda_{J^\prime J}
\rho^2 - 2 \delta_{J^\prime J} \sqrt{\frac{J}{J+1}} \rho } {1 +
\rho^2},
\end{eqnarray}

\noindent and for the $\beta$ particle longitudinal polarization

\begin{equation}
\label{eqn:Gcoeff-SM} G_{SM} = \mp 1 ,
\end{equation}

\noindent where $\delta_{J^\prime J}$ is the Kronecker delta and

\begin{equation}
\label{eqn:lambda_J'J} \lambda_{ J^\prime J} = \frac{1}{J+1}
\end{equation}

\noindent for the $J \to J^{\prime }=J$ mirror $\beta$
transitions.

%
%
%
%
%

\noindent Note that the coefficients $b_{SM} = D_{SM} = 0$ in the
standard model. When including also the effect of the Coulomb
interaction of the charged nucleus and emitted $\beta$ particle
(i.e. final state interaction, FSI) it turns out that, to first
order in $\alpha$, this depends for the $a$, $b$, $A$, $B$, $D$
and $G$ correlation coefficients on interferences between the
standard model $V,A$ coupling constants and the non-standard model
$S,T$ coupling constants \cite{jackson57}, and therefore vanishes
in the standard model. For the $N$ and $R$ correlation
coefficients, however, the final state effects contain terms that
depend on the time reversal invariant parts of the vector and/or
axial-vector coupling constants and are thus non-zero in the
standard model. To first order in $\alpha Z$ one has
\cite{jackson57}
\begin{equation}
N_{SM}^{FSI} = \mp \frac{\gamma m_e}{E_e} A_{SM}
\end{equation}
\noindent and
\begin{equation}
R_{SM}^{FSI} = \mp \frac{\alpha Z m_e}{p} A_{SM}
\end{equation}
\noindent with $E_e$ the total electron energy. Numerical
calculations \cite{vogel83} have shown that the values obtained
for $N_{FSI}$ and $R_{FSI}$ within the used approximation are
accurate at the 10\% level.

The standard model values for the coefficients $a$, $A$ and $B$ as
well as the values for $N_{FSI}$ and $R_{FSI}$ at the $\beta$
spectrum endpoint, all calculated with the  values for $\rho$
obtained from our $ft$ value analysis, are listed in
Table~\ref{correlations}. A full analysis of the sensitivity of
the different correlation coefficients to several types of physics
beyond the standard model as well as the effect of recoil order
corrections (i.e. weak magnetism) on the correlation coefficients
is in preparation and will be published elsewhere
\cite{tandecki08}.



\newpage
\begin{table}[!h]

\centering
\begin{minipage}{\textwidth}
\centering

{ \footnotesize
\begin{tabular}{ l | r@{ }c@{ }l @{$\,$} l @{  } r@{ }c@{ }l @{$\,$} l   r@{ }c@{ }l @{$\,$} l   r@{ }c@{ }ll  | r@{ }c@{ }l  c }
\hline \hline

Parent&    \multicolumn{16}{c}{Measured half-lives, $t_{1/2}$ (s) } &  \multicolumn{3}{|c}{Average half-life}   & scale        \\
\cline{2-17}
Nucleus&   \multicolumn{4}{c}{1}   &      \multicolumn{4}{c}{2}  &  \multicolumn{4}{c}{3}      &     \multicolumn{4}{c}{4}   &           \multicolumn{3}{|c}{$t_{1/2}$ (s)} &  $S$            \\
\hline
 &    &      &    &    &      &     &    &  &   &       &     &      &      &       &     &     &            &            &            &            \\

   $^{3}$H &     $4419$ &      $\pm$ &    $183$ d &     [No47] &     $4551$ &      $\pm$ &     $54$ d &     [Je50] &     $4530$ &      $\pm$ &     $27$ d &     [Jo51] &     $4479$ &      $\pm$ &     $11$ d &     [Jo55] &            &            &            &            \\

           &     $4596$ &      $\pm$ &     $66$ d &     [Po58] &     $4496$ &      $\pm$ &     $16$ d &     [Me66] &     $4474$ &      $\pm$ &     $11$ d &     [Jo67] &     $4501$ &      $\pm$ &      $9$ d &     [Ru77] &            &            &            &            \\

           &     $4498$ &      $\pm$ &     $11$ d &     [Si87] &     $4521$ &      $\pm$ &     $11$ d &     [Ol87] &     $4485$ &      $\pm$ &     $12$ d &     [Ak88] &     $4497$ &      $\pm$ &     $11$ d &     [Bu91] &            &            &            &            \\

           &     $4504$ &      $\pm$ &      $9$ d &     [Un00] &     $4500$ &      $\pm$ &      $8$ d &     [Lu00] &     $4479$ &      $\pm$ &      $7$ d &     [Ak04] &     $4497$ &      $\pm$ &      $4$ d &     [Ma06] &     $4497$ &      $\pm$ & $4$ d\footnote{We did not perform the analysis of the tritium half-lives ourselves, but rather used the value (and the references) from [Ma06]. An interesting effect is mentioned in [Ak04]; the half-life of molecular and atomic $^{3}$H would differ by about 9 days. Due to a lack of additional information on this (recently observed) effect we have not included it in the present compilation. All measurements, except for [Ak04], have been performed on molecular tritium.} &         \\

  $^{11}$C &    $20.35$ &      $\pm$ &   $0.08$ m &     [Sm41] &     $20.0$ &      $\pm$ &    $0.1$ m &     [Di51] &    $20.74$ &      $\pm$ &   $0.10$ m &     [Ku53] &    $20.26$ &      $\pm$ &   $0.10$ m &     [Ba55] &            &            &            &            \\

           &     $20.8$ &      $\pm$ &    $0.2$ m &     [Pr57] &    $20.11$ &      $\pm$ &   $0.13$ m &     [Ar58] &    $20.34$ &      $\pm$ &   $0.04$ m &    [Ka64a] &    $20.40$ &      $\pm$ &   $0.04$ m &     [Aw69] &            &            &            &            \\

           &    $20.38$ &      $\pm$ &   $0.02$ m &     [Az75] &    $20.32$ &      $\pm$ &   $0.12$ m &     [Be75] &   $20.334$ &      $\pm$ &  $0.024$ m &     [Wo02] &            &            &            &            &    $20.360$ &      $\pm$ &   $0.026$ m &        2.0 \\

  $^{13}$N &     $9.96$ &      $\pm$ &   $0.03$ m &     [Ar58] &    $9.965$ &      $\pm$ &  $0.005$ m &     [Ja60] &     $9.93$ &      $\pm$ &   $0.05$ m &     [Ki60] &    $10.05$ &      $\pm$ &   $0.05$ m &     [Bo65] &            &            &            &            \\

           &     $9.96$ &      $\pm$ &   $0.02$ m &     [Eb65] &    $9.963$ &      $\pm$ &  $0.009$ m &     [Ri68] &    $9.965$ &      $\pm$ &  $0.010$ m &     [Az77] &            &            &            &            &    $9.9647$ &      $\pm$ &  $0.0039$ m &          1 \\

  $^{15}$O &   $123.95$ &      $\pm$ &     $0.50$ &     [Pe57] &    $124.1$ &      $\pm$ &      $0.5$ &     [Ki59] &    $122.1$ &      $\pm$ &      $0.1$ &     [Ja60] &      $122.6$ &      $\pm$ &        $1.0$ &     [Ne63] &            &            &            &            \\

           &   $122.23$ &      $\pm$ &     $0.23$ &     [Az77] &            &            &            &            &            &            &            &            &            &            &            &            &    $122.24$ &      $\pm$ &      $0.27$ &        3.0 \\

  $^{17}$F &     $65.2$ &      $\pm$ &      $0.2$ &     [Wo69] &    $64.50$ &      $\pm$ &     $0.25$ &     [Al72] &    $64.31$ &      $\pm$ &     $0.09$ &     [Az77] &    $64.80$ &      $\pm$ &     $0.09$ &     [Al77] &     $64.61$ &      $\pm$ &      $0.17$ &        2.9 \\

 $^{19}$Ne &     $17.7$ &      $\pm$ &      $0.1$ &     [Pe57] &    $17.43$ &      $\pm$ &     $0.06$ &     [Ea62] &    $17.36$ &      $\pm$ &     $0.06$ &     [Go68] &    $17.36$ &      $\pm$ &     $0.06$ &     [Wi74] &            &            &            &            \\

           &   $17.219$ &      $\pm$ &    $0.017$ &     [Az75] &   $17.237$ &      $\pm$ &    $0.014$ &     [Pi85] &            &            &            &            &            &            &            &            &    $17.248$ &      $\pm$ &     $0.029$ &        2.8 \\

 $^{21}$Na &     $23.0$ &      $\pm$ &      $0.2$ &     [Ar58] &    $22.55$ &      $\pm$ &     $0.10$ &     [Al74] &    $22.47$ &      $\pm$ &     $0.03$ &     [Az75] &            &            &            &            &    $22.487$ &      $\pm$ &     $0.054$ &          1.9 \\

 $^{23}$Mg &     $12.1$ &      $\pm$ &      $0.1$ &     [Mi58] &    $11.41$ &      $\pm$ &     $0.05$ &     [Go68] &    $11.36$ &      $\pm$ &     $0.04$ &     [Al74] &    $11.26$ &      $\pm$ &     $0.08$ &     [Az74] &            &            &            &            \\

           &   $11.327$ &      $\pm$ &    $0.014$ &     [Az75] &   $11.317$ &      $\pm$ &    $0.011$ &     [Az77] &            &            &            &            &            &            &            &            &    $11.3243$ &      $\pm$ & $0.0098$\footnote{The weighted average including [Mi58] is $11.330 \pm 0.030$, compared to $11.3243 \pm 0.0098$ without [Mi58], both with scaling. Since [Mi58] has a strongly deviating value, it was decided to drop this result.} &        1.2 \\

 $^{25}$Al &     $7.24$ &      $\pm$ &     $0.03$ &     [Mu58] &     $7.23$ &      $\pm$ &     $0.02$ &     [Ju71] &    $7.177$ &      $\pm$ &    $0.023$ &     [Ta73] &    $7.174$ &      $\pm$ &    $0.007$ &     [Az75] &    $7.182$ &      $\pm$ &    $0.012$ &        1.9 \\

 $^{27}$Si &     $4.14$ &      $\pm$ &     $0.03$ &     [Mi58] &     $4.16$ &      $\pm$ &     $0.03$ &     [Su62] &     $4.19$ &      $\pm$ &     $0.02$ &     [Bl68] &     $4.17$ &      $\pm$ &     $0.01$ &     [Go68] &            &            &            &            \\

           &     $4.21$ &      $\pm$ &     $0.03$ &     [Gr71] &    $4.109$ &      $\pm$ &    $0.004$ &     [Az75] &    $4.206$ &      $\pm$ &    $0.008$ &     [Ge76] &     $4.09$ &      $\pm$ &     $0.02$ &     [Ba77] &     $4.135$ &      $\pm$ &     $0.019$ &        6.0 \\

  $^{29}$P &     $4.19$ &      $\pm$ &     $0.02$ &     [Ja60] &     $4.15$ &      $\pm$ &     $0.03$ &     [Sc70] &    $4.149$ &      $\pm$ &    $0.005$ &     [Ta73] &    $4.083$ &      $\pm$ &     $0.012$ &     [Az75] &            &            &            &            \\

           &    $4.084$ &      $\pm$ &    $0.022$ &     [Wi80] &            &            &            &            &            &            &            &            &            &            &            &            &     $4.140$ &      $\pm$ &     $0.016$ &        3.6 \\

  $^{31}$S &     $2.66$ &      $\pm$ &     $0.03$ &     [Ha52] &     $2.40$ &      $\pm$ &     $0.07$ &     [Hu54] &     $2.80$ &      $\pm$ &     $0.05$ &     [Cl58] &     $2.72$ &      $\pm$ &     $0.02$ &     [Mi58] &            &            &            &            \\

           &     $2.57$ &      $\pm$ &     $0.01$ &     [Ja60] &     $2.61$ &      $\pm$ &     $0.05$ &     [Li60] &     $2.58$ &      $\pm$ &     $0.06$ &     [Wa60] &    $2.605$ &      $\pm$ &    $0.012$ &     [Al74] &            &            &            &            \\

           &    $2.543$ &      $\pm$ &    $0.008$ &     [Az77] &    $2.562$ &      $\pm$ &    $0.007$ &     [Wi80] &            &            &            &            &            &            &            &            &     $2.574$ &      $\pm$ & $0.017$\footnote{Note that without [Mi58], the central value of which differs from later results, the weighted average becomes $2.567 \pm 0.011$ s. } &        4.2\\

 $^{33}$Cl &     $2.53$ &      $\pm$ &     $0.02$ &     [Mu58] &     $2.51$ &      $\pm$ &     $0.02$ &     [Ja60] &     $2.47$ &      $\pm$ &     $0.02$ &     [Sc70] &    $2.513$ &      $\pm$ &    $0.004$ &     [Ta73] &            &            &            &            \\

           &    $2.507$ &      $\pm$ &    $0.008$ &     [Az77] &            &            &            &            &            &            &            &            &            &            &            &            &    $2.5111$ &      $\pm$ &    $0.0040$ &        1.2 \\

 $^{35}$Ar &     $1.79$ &      $\pm$ &     $0.01$ &     [Ja60] &    $1.770$ &      $\pm$ &    $0.006$ &     [Wi69] &    $1.774$ &      $\pm$ &    $0.003$ &     [Az77] &   $1.7754$ &      $\pm$ &   $0.0011$ &     [Ia06] &   $1.7752$ &      $\pm$ &   $0.0010$ &          1 \\

  $^{37}$K &     $1.23$ &      $\pm$ &     $0.02$ &     [Sc58] &     $1.25$ &      $\pm$ &     $0.04$ &     [Ka64] &    $1.223$ &      $\pm$ &    $0.008$ &     [Az77] &            &            &            &            &    $1.2248$ &      $\pm$ &    $0.0073$ &          1 \\

 $^{39}$Ca &     $0.90$ &      $\pm$ &     $0.01$ &     [Kl54] &    $0.876$ &      $\pm$ &    $0.012$ &     [Cl58] &    $0.860$ &      $\pm$ &    $0.005$ &     [Mi58] &    $0.873$ &      $\pm$ &    $0.008$ &     [Li60] &            &            &            &            \\

           &    $0.865$ &     $^+_-$ & $^{0.007}_{0.017}$ &     [Ka68]$^e$ &   $0.8604$ &      $\pm$ &   $0.0030$ &     [Al73] &   $0.8594$ &      $\pm$ &   $0.0016$ &     [Az77] &            &            &            &            &    $0.8609$ &      $\pm$ &    $0.0028$ &        2.2 \\

 $^{41}$Sc &    $0.628$ &      $\pm$ &    $0.014$ &     [Ja60] &    $0.596$ &      $\pm$ &    $0.006$ &     [Yo65] &   $0.5963$ &      $\pm$ &   $0.0017$ &     [Al73] &    $0.591$ &      $\pm$ &    $0.005$ &     [Ta73] &   $0.5962$ &      $\pm$ &   $0.0022$ &          1.4 \\

 $^{43}$Ti &     $0.58$ &      $\pm$ &     $0.04$ &     [Sc48] &    $0.528$ &      $\pm$ &    $0.003$ &     [Ja60] &     $0.56$ &      $\pm$ &     $0.02$ &     [Ja61] &     $0.50$ &      $\pm$ &     $0.02$ &     [Pl62] &            &            &            &            \\

           &     $0.40$ &      $\pm$ &     $0.05$ &     [Va63] &     $0.49$ &      $\pm$ &     $0.01$ &     [Al67] &   $0.54$ &      $\pm$ &     $0.01$ &     [Va69] &    $0.509$ &      $\pm$ &    $0.005$ &     [Ho87] &    $0.5222$ &      $\pm$ & $0.0057$\footnote{The weighted average discarding [Ja60] is $0.5124 \pm 0.0085$ s, compared to $0.5222 \pm 0.0057$ s, both with scaling included. Since there is no clear reason to drop [Ja60] it was decided to keep it. Note that this is the most precise result, yet it dates from $1960$. } &        2.4 \\

  $^{45}$V &    $0.539$ &      $\pm$ &    $0.018$ &     [Ho82] &   $0.5472$ &      $\pm$ &   $0.0053$ &     [Ha87] &            &            &            &            &            &            &            &            &    $0.5465$ &      $\pm$ &    $0.0051$ &          1 \\

 $^{47}$Cr &   $0.4600$ &      $\pm$ &   $0.0015$ &     [Ed77] &    $0.508$ &      $\pm$ &    $0.010$ &     [Bu85] &   $0.4720$ &      $\pm$ &   $0.0063$ &     [Ha87] &            &            &            &            &    $0.4616$ &      $\pm$ &    $0.0051$ &        3.6 \\

 $^{49}$Mn &    $0.384$ &      $\pm$ &    $0.017$ &     [Ha80] &   $0.3817$ &      $\pm$ &   $0.0074$ &     [Ha87] &            &            &            &            &            &            &            &            &    $0.3821$ &      $\pm$ &    $0.0068$ &          1 \\

 $^{51}$Fe &    $0.310$ &      $\pm$ &    $0.005$ &     [Ay84] &   $0.3050$ &      $\pm$ &   $0.0043$ &     [Ha87] &            &            &            &            &            &            &            &            &     $0.3071$ &      $\pm$ &     $0.0033$ &        1 \\

 $^{53}$Co &    $0.262$ &      $\pm$ &    $0.025$ &     [Ko73] &    $0.240$ &      $\pm$ &    $0.025$ &     [Ho89] &    $0.267$ &      $\pm$ &    $0.025$ &     [Ha87] &    $0.240$ &      $\pm$ &    $0.009$ &     [Lo02] &    $0.2446$ &      $\pm$ &    $0.0076$ &          1 \\

 $^{55}$Ni &    $0.189$ &      $\pm$ &    $0.005$ &     [Ho77] &    $0.208$ &      $\pm$ &    $0.005$ &     [Ay84] &   $0.2121$ &      $\pm$ &   $0.0038$ &     [Ha87] &    $0.204$ &      $\pm$ &    $0.003$ &     [Re99] &            &            &            &            \\

           &    $0.196$ &      $\pm$ &    $0.005$ &     [Lo02] &            &            &            &            &            &            &            &            &            &            &            &            &    $0.2033$ &      $\pm$ &    $0.0037$ &          2.0 \\

 $^{57}$Cu &   $0.1994$ &      $\pm$ &   $0.0032$ &     [Sh89] &   $0.1963$ &      $\pm$ &   $0.0007$ &     [Se96] &            &            &            &            &            &            &            &            &   $0.19644$ &      $\pm$ &   $0.00068$ &          1 \\

 $^{59}$Zn &   $0.1820$ &      $\pm$ &   $0.0018$ &     [Ar84] &    $0.173$ &      $\pm$ &    $0.014$ &     [Lo02] &            &            &            &            &            &            &            &            &   $0.1819$ &      $\pm$ &   $0.0018$ &          1 \\

 $^{61}$Ga &     $0.15$ &      $\pm$ &     $0.03$ &     [Wi93] &    $0.168$ &      $\pm$ &    $0.003$ &     [We02] &    $0.148$ &      $\pm$ &    $0.019$ &     [Lo02] &            &            &            &            &    $0.1673$ &      $\pm$ &    $0.0029$ &          1 \\

 $^{63}$Ge &    $0.095$ &     $^+_-$ & $^{0.023}_{0.025}$ &     [Wi93]$^e$ &    $0.095$ &     $^+_-$ & $^{0.023}_{0.020}$ &     [Sh93]$^e$ &    $0.150$ &      $\pm$ &    $0.009$ &     [Lo02] &            &            &            &            &    $0.137$ &      $\pm$ &    $0.016$ &        2.1 \\

 $^{65}$As &     $0.19$ &     $^+_-$ & $^{0.11}_{0.07}$ &     [Wi93]$^e$ &    $0.190$ &      $\pm$ &    $0.011$ &     [Mo95] &    $0.126$ &      $\pm$ &    $0.016$ &     [Lo02] &            &            &            &            &     $0.170$ &      $\pm$ &     $0.030$ &        3.3 \\

 $^{67}$Se &    $0.107$ &      $\pm$ &    $0.035$ &     [Ba94] &    $0.060$ &     $^+_-$ & $^{0.017}_{0.011}$ &     [Bl95]$^e$ &    $0.136$ &      $\pm$ &    $0.012$ &     [Lo02] &            &            &            &            &    $0.106$ &      $\pm$ &    $0.024$ &        2.7 \\

 $^{71}$Kr &   $ 0.097$ &      $\pm$ &    $0.009$ &     [Ew81] &    $0.064$ &     $^+_-$ & $^{0.008}_{0.005}$ &     [Bl95]$^e$ &    $0.100$ &      $\pm$ &    $0.003$ &     [Oi97] &            &            &            &            &    $0.0944$ &      $\pm$ &    $0.0086$ &        3.3 \\

 $^{75}$Sr &   $ 0.088$ &      $\pm$ &    $0.003$ &     [Hu03] &            &            &            &            &            &            &            &            &            &            &            &            &    $0.088$ &      $\pm$ &    $0.003$ &           \\

  $^{77}$Y &    $0.057$ &     $^+_-$ & $^{0.022}_{0.012}$ &     [Ki01]$^e$ &            &            &            &            &            &            &            &            &            &            &            &            &    $0.065$ &      $\pm$ &    $0.017$ &           \\

 $^{79}$Zr &   $ 0.056$ &      $\pm$ &    $0.030$ &     [Bl99] &   &       &    &      &            &            &            &            &            &            &            &            &    $0.056$ &      $\pm$ &    $0.030$ &         \\

 $^{83}$Mo &    $0.006$ &     $^+_-$ & $^{0.030}_{0.003}$ &     [Ki01]$^e$ &            &            &            &            &            &            &            &            &            &            &            &            &    $0.028$ &      $\pm$ &    $0.019$ &           \\

\hline \hline

\end{tabular}

\caption{  Half-lives, $t_{1/2}$, of the mirror nuclei, expressed
in seconds unless specified differently (days (d), minutes (m)).
References to data listed in this table are given in
Table~\ref{usedref_half-life}. References to data that were not
used are listed in Table~\ref{unusedref_half-life} together with
the reason for their rejection. The scale factor $S$ listed in the
last column is defined in Eq.~\ref{scale}.}

\footnotetext[5]{These asymmetric errors have been symmetrized for
the analysis by using standard recommendations of the Particle
Data Group.}

}

\end{minipage}

\label{halflifes}

\end{table}




\newpage
\begin{table}[!h]

\centering
\begin{minipage}{\textwidth}
\centering

{ \small
\begin{tabular}{ l | r@{ }c@{ }l @{$\,$} l @{  } r@{ }c@{ }l @{$\,$} l   r@{ }c@{ }l @{$\,$} l     | r@{ }c@{ }l  c }
\hline \hline
Parent  &    \multicolumn{12}{c|}{Measured branching ratio, BR (\%)} &  \multicolumn{3}{c}{Average value}   & scale        \\
nucleus    &   \multicolumn{4}{c}{1}    &           \multicolumn{4}{c}{2}             &           \multicolumn{4}{c|}{3}          &    \multicolumn{3}{c}{$BR$  $(\%)$}   & $S$      \\

\hline
    &    &            &            &     &            &            &            &            &            &            &            &            &     &            &      &      \\
   $^{3}$H &      $100$ &            &            &     [Ti87] &            &            &            &            &            &            &            &            &      $100$ &            &    &         \\

  $^{11}$C &      $100$ &            &            &     [Aj75] &            &            &            &            &            &            &            &            &      $100$ &            &     &        \\

  $^{13}$N &      $100$ &            &            &     [Aj70] &            &            &            &            &            &            &            &            &      $100$ &            &      &       \\

  $^{15}$O &      $100$ &            &            &     [Aj70] &            &            &            &            &            &            &            &            &      $100$ &            &     &        \\

  $^{17}$F &      $100$ &            &            &     [Aj70] &            &            &            &            &            &            &            &            &      $100$ &            &      &       \\

 $^{19}$Ne & BR(1.55MeV): &            &            &            &   $0.0021$ &      $\pm$ &   $0.0003$ &     [Al76] &   $0.0023$ &      $\pm$ &   $0.0003$ &     [Ad83] &            &            &    &         \\

           & BR(0.11MeV): &            &            &            &    $0.012$ &      $\pm$ &    $0.002$ &     [Ad81] &    $0.011$ &      $\pm$ &    $0.009$ &     [Sa93] &   $99.9858$ &      $\pm$ &    $0.0020$  & 1 \\

 $^{21}$Na &     $94.9$ &      $\pm$ &      $0.2$ &     [Al74] &     $95.8$ &      $\pm$ &      $0.2$ &     [Az77] &    $94.98$ &      $\pm$ &     $0.13$ &     [Wi80] &            &            &          &    \\

           &    $95.26$ &      $\pm$ &     $0.04$ &     [Ia06] &     $95.15$       &  $\pm$          &   $0.12$         & [Ac07]           &            &            &            &            &    $95.235$ &      $\pm$ &     $0.069$  & 2.0 \\

 $^{23}$Mg &     $90.9$ &      $\pm$ &      $0.5$ &     [Ta60] &     $91.4$ &      $\pm$ &      $0.4$ &    [Go68a] &     $90.9$ &      $\pm$ &      $0.4$ &     [Al74] &            &            &          &   \\

           &     $91.9$ &      $\pm$ &      $0.4$ &     [Ma74] &     $92.2$ &      $\pm$ &      $0.2$ &     [Az77] &            &            &            &            &    $91.78$ &      $\pm$ &     $0.26$  & 1.8\\

 $^{25}$Al &    $99.16$ &      $\pm$ &     $0.07$ &     [Ju71] &     $99.1$ &      $\pm$ &      $0.2$ &     [Ma69] &    $99.11$ &      $\pm$ &     $0.08$ &     [Ma76] &            &            &           &   \\

           &    $99.16$ &      $\pm$ &     $0.04$ &     [Az77] &            &            &            &            &            &            &            &            &    $99.151$ &      $\pm$ &     $0.031$  & 1 \\

 $^{27}$Si &    $99.90$ &      $\pm$ &     $0.02$ &     [Go64] &    $99.80$ &      $\pm$ &     $0.07$ &     [De71] &    $99.82$ &      $\pm$ &     $0.05$ &     [Be71] &            &            &            &   \\

           &    $99.77$ &      $\pm$ &     $0.02$ &     [Ma74] &    $99.81$ &      $\pm$ &     $0.01$ &     [Az77] &            &            &            &            &    $99.818$ &      $\pm$ &     $0.022$  & 2.8 \\

  $^{29}$P &     $98.4$ &      $\pm$ &      $0.3$ &     [Lo62] &    $98.11$ &      $\pm$ &     $0.30$ &     [Az77] &    $98.29$ &      $\pm$ &     $0.03$ &     [Wi80] &    $98.290$ &      $\pm$ &     $0.030$  & 1\\

  $^{31}$S &     $98.9$ &      $\pm$ &      $0.1$ &     [Ta60] &     $99.2$ &      $\pm$ &      $0.4$ &     [De71] &    $98.75$ &      $\pm$ &     $0.06$ &     [Al74] &            &            &            &  \\

           &    $98.89$ &      $\pm$ &     $0.20$ &     [Az77] &    $98.86$ &      $\pm$ &     $0.04$ &     [Wi80] &            &            &            &            &    $98.837$ &      $\pm$ &     $0.031$  & 1\\

 $^{33}$Cl &     $98.3$ &      $\pm$ &      $0.2$ &     [Ba70] &    $98.58$ &      $\pm$ &     $0.19$ &     [Wi80] &            &            &            &            &    $98.45$ &      $\pm$ &     $0.14$ & 1 \\

 $^{35}$Ar &    $98.32$ &      $\pm$ &     $0.07$ &     [Wi69] &    $98.55$ &      $\pm$ &     $0.05$ &     [De71] &     $98.3$ &      $\pm$ &      $0.2$ &     [Ge71] &            &            &            &  \\

           &     $98.0$ &      $\pm$ &      $0.2$ &     [Az77] &    $98.24$ &      $\pm$ &     $0.05$ &     [Wi80] &    $98.24$ &      $\pm$ &     $0.10$ &     [Ad84] &    $98.358$ &      $\pm$ &     $0.066$  & 2.2\\

  $^{37}$K &     $98.0$ &      $\pm$ &      $0.4$ &     [Ka64] &     $98.5$ &      $\pm$ &      $0.2$ &     [Ma76] &     $97.8$ &      $\pm$ &      $0.2$ &     [Az77] &     &      &       & \\

           &    $97.89$ &      $\pm$ &     $0.11$ &     [Ha97] &            &            &            &            &            &            &            &            &     $97.99$       &   $\pm$ &     $0.14$      & 1.7 \\

 $^{39}$Ca &  $99.9975$ &      $\pm$ &   $0.0002$ &     [Ha94] &            &            &            &            &            &            &            &            &  $99.9975$ &      $\pm$ &   $0.0002$ &  \\

 $^{41}$Sc &   $99.963$ &      $\pm$ &    $0.003$ &     [Wi80] &            &            &            &            &            &            &            &            &   $99.963$ &      $\pm$ &    $0.003$  &  \\

 $^{43}$Ti &     $90.2$ &      $\pm$ &      $0.8$ &     [Ho87] &            &            &            &            &            &            &            &            &     $90.2$ &      $\pm$ &      $0.8$  &  \\

  $^{45}$V &     $95.7$ &      $\pm$ &      $1.5$ &     [Ho82] &            &            &            &            &            &            &            &            &     $95.7$ &      $\pm$ &      $1.5$  &  \\

 $^{47}$Cr &     $96.3$ &      $\pm$ &      $1.2$ &     [Bu85] &            &            &            &            &            &            &            &            &     $96.3$ &      $\pm$ &      $1.2$ &  \\

 $^{49}$Mn &     $93.6$ &      $\pm$ &      $2.6$ &     [Ha80] &     $91.9$ &      $\pm$ &      $2.8$ &     [Ho89] &            &            &            &            &     $92.8$ &      $\pm$ &      $1.9$ & 1 \\

 $^{51}$Fe &     $95.0$ &      $\pm$ &      $1.3$ &     [Ay84] &     $93.8$ &      $\pm$ &      $1.3$ &     [Ho89] &            &            &            &            &     $94.40$ &      $\pm$ &      $0.92$ & 1 \\

 $^{53}$Co &     $94.4$ &      $\pm$ &      $1.7$ &     [Ho89] &            &            &            &            &            &            &            &            &     $94.4$ &      $\pm$ &      $1.7$ &  \\

 $^{57}$Cu &     $89.9$ &      $\pm$ &      $0.8$ &     [Se96] &            &            &            &            &            &            &            &            &     $89.9$ &      $\pm$ &      $0.8$ &  \\

 $^{59}$Zn &     $93.0$ &      $\pm$ &      $3.0$ &     [Ho81] &     $94.1$ &      $\pm$ &      $0.8$ &     [Ar84] &            &            &            &            &     $94.03$ &      $\pm$ &      $0.77$ & 1 \\

 $^{61}$Ga &       $94$ &      $\pm$ &        $1$ &     [We02] &            &            &            &            &            &            &            &            &       $94$ &      $\pm$ &        $1$ &  \\

 $^{71}$Kr &     $82.1$ &      $\pm$ &      $1.6$ &     [Oi97] &            &            &            &            &            &            &            &            &     $82.1$ &      $\pm$ &      $1.6$ &  \\

 $^{75}$Sr &     $90.3$ &     $^+_-$ & $^{1.9}_{2.8}$ & [Hu03]\footnote{These asymmetric errors have been symmetrized for the analysis by using standard recommendations of the Particle Data Group.} &            &            &            &            &            &            &            &            &     $89.6$ &      $\pm$ &      $2.4$ &  \\

\hline \hline

\end{tabular}

\caption{  Branching ratios, $\BR$, for the $T = 1/2$  mirror
$\beta$ transitions. References to data listed here are given in
Table \ref{usedref_BR}. References to rejected are listed in Table
\ref{unusedref_BR}. \label{branchings} }

}
\end{minipage}

\end{table}



\newpage
\begin{table}[!h]

\centering

\begin{tabular}{ l   r@{ }c@{ }l@{ }   c@{\extracolsep{\fill} }   r@{ }c@{ }l@{ }   r@{ }c@{ }l@{ }   r@{ }c@{ }l@{ }   }
\hline \hline 
Parent  &    \multicolumn{3}{c}{$t_{1/2}$} &  {$P_{EC}$}  &  \multicolumn{3}{c}{$\br$} &  \multicolumn{3}{c}{$t$ } &  \multicolumn{3}{c}{$Q_{EC}$}            \\
nucleus  &    \multicolumn{3}{c}{(s)} &       {$(\%)$}   &  \multicolumn{3}{c}{$(\%)$} &  \multicolumn{3}{c}{(s)} &  \multicolumn{3}{c}{(keV)}            \\
    \hline
   &      &               &   &  &  &   &   &     \\
   $^{3}$H &  $(38854$ &      $\pm$ & $ 35 ) \times 10^{4}$ &   N/A  &     \  &      $100$ &        \ & $(38854 $ &      $\pm$ & $35 ) \times 10^4$ &  $18.5912$ &      $\pm$ &    $0.0010$ \\

  $^{11}$C &   $1221.6$ &      $\pm$ &     $1.5$  &     $0.231$ &               \  &    $100$ &        \ &   $1224.4$ &      $\pm$ &      $1.5$ &  $1982.40$ &      $\pm$ &      $0.90$ \\

  $^{13}$N &   $597.88$ &      $\pm$ &    $0.23$  &     $0.196$ &               \ &     $100$ &        \ &   $599.05$ &      $\pm$ &     $0.23$ &  $2220.47$ &      $\pm$ &     $0.27$ \\

  $^{15}$O &    $122.24$ &      $\pm$ &      $0.27$ &     $0.100$ &               \ &     $100$ &        \ &    $122.37$ &      $\pm$ &      $0.27$ &  $2754.16$ &      $\pm$ &      $0.50$ \\

  $^{17}$F &     $64.61$ &      $\pm$ &      $0.17$ &     $0.147$ &                 \ &   $100$ &        \ &     $64.70$ &      $\pm$ &      $0.17$ &  $2760.51$ &      $\pm$ &     $0.27$ \\

 $^{19}$Ne &    $17.248$ &      $\pm$ &     $0.029$ &     $0.101$ &  $99.9858$ \  &      $\pm$ &  $0.0020$ \ &   $17.268$ &      $\pm$ &    $0.029$ &  $3238.83$ &      $\pm$ &      $0.30$ \\

 $^{21}$Na &    $22.487$ &      $\pm$ &     $0.054$ &    $0.094$ &   $95.235$ \  &      $\pm$ &   $0.069$ \ &    $23.634$ &      $\pm$ &     $0.060$ &  $3547.58$ &      $\pm$ &      $0.70$ \\

 $^{23}$Mg &   $11.3243$ &      $\pm$ &    $0.0098$ &    $0.073$ &   $91.78$ \  &      $\pm$ &   $0.26$ \ &    $12.348$ &      $\pm$ &     $0.037$ &  $4056.1$ &      $\pm$ &      $1.3$ \\

 $^{25}$Al &    $7.182$ &      $\pm$ &    $0.012$ &    $0.079$ &   $99.151$ \  &      $\pm$ &   $0.031$ \ &    $7.250$ &      $\pm$ &    $0.012$ &  $4276.63$ &      $\pm$ &      $0.50$ \\

 $^{27}$Si &     $4.135$ &      $\pm$ &     $0.019$ &    $0.065$ &   $99.818$ \  &      $\pm$ &   $0.022$ \ &     $4.145$ &      $\pm$ &     $0.020$ &  $4812.36$ &      $\pm$ &      $0.10$ \\

  $^{29}$P &     $4.140$ &      $\pm$ &     $0.016$ &    $0.075$ &   $98.290$ \  &      $\pm$ & $0.030$ \ &     $4.215$ &      $\pm$ &     $0.016$ &  $4942.45$ &      $\pm$ &      $0.60$ \\

  $^{31}$S &     $2.574$ &      $\pm$ &     $0.017$ &    $0.069$ &   $98.837$ \  &      $\pm$ & $0.031$ \ &    $2.606$ &      $\pm$ &    $0.017$ &   $5396.3$ &      $\pm$ &      $1.5$ \\

 $^{33}$Cl &    $2.5111$ &      $\pm$ &    $0.0040$ &    $0.075$ &   $98.45$ \  &      $\pm$ & $0.14$ \ &   $2.5526$ &      $\pm$ &   $0.0055$ &  $5582.59$ &      $\pm$ &      $0.40$ \\

 $^{35}$Ar &   $1.7752$ &      $\pm$ &   $0.0010$ &    $0.073$ &   $98.358$ \  &      $\pm$ & $0.066$ \ &   $1.8062$ &      $\pm$ &   $0.0016$ &  $5966.14$ &      $\pm$ &      $0.70$ \\

  $^{37}$K &    $1.2248$ &      $\pm$ &    $0.0073$ &    $0.080$ &   $97.99$ \  &      $\pm$ & $0.14$ \ &    $1.2510$ &      $\pm$ &   $0.0077$ &  $6147.46$ &      $\pm$ &      $0.20$ \\

 $^{39}$Ca &    $0.8609$ &      $\pm$ &    $0.0028$ &    $0.078$ &   $99.9975$ \  &      $\pm$ & $0.0002$ \ &   $0.8616$ &      $\pm$ &   $0.0028$ &   $6532.61$ &      $\pm$ &      $1.9$ \\

 $^{41}$Sc &   $0.5962$ &      $\pm$ &   $0.0022$ &    $0.096$ &   $99.963$ \  &      $\pm$ & $0.003$ \ &   $0.5970$ &      $\pm$ &   $0.0022$ &  $6495.37$ &      $\pm$ &     $0.16$ \\

 $^{43}$Ti &    $0.5222$ &      $\pm$ &    $0.0057$ &    $0.094$ &   $90.2$ \  &      $\pm$ & $0.8$ \ &   $0.5795$ &      $\pm$ &   $0.0082$ &   $6866.9$ &      $\pm$ &      $7.3$ \\

  $^{45}$V &    $0.5465$ &      $\pm$ &    $0.0051$ &    $0.098$ &   $95.7$ \  &      $\pm$ &    $1.5$ \ &    $0.572$ &      $\pm$ &    $0.010$ &     $7126$ &      $\pm$ &       $17$ \\

  \hline
  \hline

\end{tabular}
\caption{   Overview of the adopted values for the half-lifes,
$t_{1/2}$, and the branching ratios, $\br$, for the $T = 1/2$
mirror $\beta$ transitions, together with the electron capture
probabilities, $P_{EC}$, (from \cite{bambynek77,firestone96}), the
deduced partial half-lifes, $t$, (cf. Eq.~(\ref{eq:tECBR} )) and
the $Q_{EC}$ values (from \cite{audi03} ).} \label{adopted_t_BR}

\end{table}




\newpage
\begin{sidewaystable}[!h]

\centering
\begin{tabular*}{0.99\textwidth}{ l   r@{\extracolsep{\fill} }c@{ \extracolsep{\fill}} l@{\extracolsep{\fill} }  c@{\extracolsep{\fill} } c@{\extracolsep{\fill} } c@{\extracolsep{\fill} }   c@{\extracolsep{\fill} }   c@{ \extracolsep{\fill}}  c@{ \extracolsep{\fill}}  c@{ \extracolsep{\fill}} c@{ \extracolsep{\fill}}   c@{ \extracolsep{\fill}}  c@{ \extracolsep{\fill}}  c@{ \extracolsep{\fill}}}
\hline \hline
Parent  &    \multicolumn{3}{c}{$f_V$} &  {$f_V t$} &  $\frac{f_A}{f_V}$ &  $\delta_1$ &  $\delta_2$ &  $\delta_3$  & $\delta_{\alpha^2}$  &  {$\delta_R'$} & {$\delta^V_{c1}$} & {$\delta^V_{c2}$}   & {$\delta^V_{NS}$ }  & {$\delta^V_C - \delta^V_{NS}$ }     \\
nucleus  &    & &    &  $(s)$   &              &  $(\%)$  &    $(\%)$   &      $(\%)$  & $(\%)$ &  {$(\%)$ }    &  {$(\%)$ }             &     {$(\%)$ } &  $(\%)$  & $(\%)$                              \\
\hline
&         &   &   &     &     &     &      &      &      &      &      &  \\
   $^{3}$H &  $(2.8757$ &      $\pm$ & $0.0026 ) \times 10^{-6}$ &  $1117.3(14)$ & $1.00492$ &    $1.816$ &   $-0.084$ &    $0.001$ & $0.035$  &    $1.768(1)$ &    $0.002(2)$ &    $0.025(1)$ &    $-0.13(2)$ &    $0.16(2)$ \\

  $^{11}$C &    $3.193$ &      $\pm$ &    $0.012$ &  $3910(16)$ & $1.01052$ &    $1.450$ &    $0.179$ &    $0.004$ &    $0.027$ &   $1.660(4)$ &    $0.003(3)$ &    $0.925(20)$ &    $-0.12(2)$ &    $1.04(3)$ \\

  $^{13}$N &    $7.716$ &      $\pm$ &    $0.007$ &  $4622.0(47)$ &  $1.00450$ &    $1.396$ &    $0.208$ &    $0.006$ &    $0.025$ &   $1.635(6)$ &    $0.006(6)$&    $0.265(15)$ &    $-0.06(2)$ &    $0.33(3)$ \\

  $^{15}$O &   $35.500$ &      $\pm$ &    $0.044$ &  $4344(11)$ & $1.00263$ &    $1.298$ &    $0.225$ &    $0.008$ &    $0.024$ &   $1.555(8)$ &    $0.016(10)$ &    $0.165(15)$ &    $-0.04(2)$ &    $0.22(3)$ \\

  $^{17}$F &   $35.217$ &      $\pm$ &    $0.024$ &  $2278.6(61)$ & $1.01704$ &    $1.297$ &    $0.257$ &    $0.010$ &    $0.023$ &   $1.587(10)$ &    $0.025(10)$ &    $0.560(25)$ &    $-0.04(2)$ &    $0.62(3)$ \\

 $^{19}$Ne &   $98.532$ &      $\pm$ &    $0.058$ &  $1701.4(30)$ & $1.01428$ &    $1.226$ &    $0.272$ &    $0.012$ &    $0.022$ &   $1.533(12)$ &     $0.140(30)$ &    $0.275(25)$ &    $-0.11(2)$ &    $0.52(4)$ \\

 $^{21}$Na &   $170.97$ &      $\pm$ &     $0.21$ &  $4041(11)$ & $1.01801$ &    $1.186$ &    $0.291$ &    $0.015$ &    $0.021$ &   $1.514(15)$ &    $0.028(10)$ &    $0.320(25)$ &    $-0.06(2)$ &    $0.41(3)$ \\

 $^{23}$Mg &   $378.59$ &      $\pm$ &     $0.73$ &  $4675(17)$ & $1.01935$ &    $1.129$ &    $0.309$ &    $0.017$ &    $0.020$ &   $1.476(17)$ &    $0.023(10)$ &    $0.270(20)$ &    $-0.11(2)$ &    $0.40(3)$ \\

 $^{25}$Al &   $508.45$ &      $\pm$ &     $0.35$ &  $3686.1(67)$ & $1.02373$ &    $1.108$ &    $0.328$ &    $0.020$ &    $0.020$ &   $1.475(20)$ &    $0.061(40)$ &    $0.400(25)$ &    $-0.06(2)$ &    $0.52(5)$ \\

 $^{27}$Si &   $993.61$ &      $\pm$ &     $0.12$ &  $4119(19)$ & $1.02697$ &    $1.059$ &    $0.342$ &    $0.023$ &    $0.019$ &   $1.443(23)$ &    $0.052(30)$ &    $0.260(15)$ &    $-0.11(2)$ &    $0.42(4)$ \\

  $^{29}$P &   $1136.7$ &      $\pm$ &      $0.8$ &  $4791(18)$ & $1.02231$ &    $1.047$ &    $0.361$ &    $0.026$ &    $0.020$ &   $1.453(26)$ &    $0.091(40)$ &    $0.885(35)$ &    $-0.09(2)$ &    $1.07(6)$ \\

  $^{31}$S &   $1841.5$ &      $\pm$ &      $2.9$ &  $4798(33)$ & $1.01951$ &    $1.011$ &    $0.372$ &    $0.029$ &    $0.018$ &   $1.430(29)$ &     $0.220(30)$ &    $0.495(20)$ &    $-0.08(2)$ &    $0.79(4)$ \\

 $^{33}$Cl &   $2190.0$ &      $\pm$ &      $0.9$ &  $5590(12)$ & $0.98777$ &    $0.996$ &    $0.389$ &    $0.032$ &    $0.018$ &   $1.435(32)$ &    $0.145(20)$ &    $0.720(55)$ &    $-0.06(2)$ &    $0.93(6)$ \\

 $^{35}$Ar &   $3121.9$ &      $\pm$ &      $2.1$ &  $5638.8(63)$ & $0.98938$ &    $0.969$ &    $0.399$ &    $0.035$ &    $0.017$ &   $1.421(35)$ &    $0.038(10)$ &    $0.455(45)$ &    $-0.04(2)$ &    $0.53(5)$ \\

  $^{37}$K &   $3623.9$ &      $\pm$ &      $0.7$ &  $4533(28)$ & $1.00456$ &    $0.958$ &    $0.417$ &    $0.039$ &    $0.017$ &   $1.431(39)$ &    $0.054(10)$ &    $0.680(60)$ &    $-0.06(2)$ &    $0.79(6)$ \\

 $^{39}$Ca &   $4985.8$ &      $\pm$ &      $8.0$ &  $4296(16)$ & $1.00101$ &    $0.934$ &    $0.428$ &    $0.042$ &    $0.017$ &   $1.421(42)$ &     $0.330(60$) &    $0.525(55)$ &    $-0.09(2)$ &    $0.95(8)$ \\

 $^{41}$Sc &   $4745.0$ &      $\pm$ &      $0.6$ &  $2833(11)$ & $1.03671$ &    $0.941$ &    $0.449$ &    $0.047$ &    $0.017$ &   $1.453(47)$ &    $0.041(20)$ &    $0.780(60)$ &    $-0.04(2)$ &    $0.86(7)$ \\

 $^{43}$Ti &     $6336$ &      $\pm$ &       $37$ &  $3671(56)$ & $1.03184$ &    $0.918$ &    $0.459$ &    $0.050$ &    $0.016$ &   $1.444(50)$ &     $0.170(100)$ &    $0.330(30)$ &    $-0.13(2)$ &    $0.63(11)$ \\

  $^{45}$V &     $7628$ &      $\pm$ &      $100$ &  $4361(98)$ & $1.04112$ &    $0.903$ &    $0.466$ &    $0.054$ &    $0.016$ &   $1.439(54)$ &     $0.170(100)$ &    $0.695(70)$ &    $-0.06(2)$ &    $0.93(12)$ \\

\hline \hline
\end{tabular*}
\caption{   Calculated quantities and corrections needed to obtain
the $\ft^{\mbox{mirror}}$ values (Eq.~(23)). Details are given in
the text.} \label{calculated quantities}

\end{sidewaystable}



\newpage
\begin{table}[!h]

\centering
\begin{tabular}{ l   r@{ }c@{ }p{1cm }      p{1.1cm}  r@{ }c@{ }p{1cm }   p{1.1cm}     }

\hline \hline
Parent  &    \multicolumn{3}{c}{$\ft$} &  $\delta \ft$ &  \multicolumn{3}{c}{$\rho $}   &  $\delta\rho$       \\
nucleus  &   \multicolumn{3}{c}{(s)}             &    (\%)       &                     & &       &      ($\%$)               \\

\hline

 &  &&  &  &  && &  \\

   $^{3}$H &   $1135.3$ &      $\pm$ &      $1.5$ &     $0.13$ &  $-2.0951$ &      $\pm$ &   $0.0020$ &     $0.10$ \\

  $^{11}$C &     $3933$ &      $\pm$ &       $16$ &     $0.41$ &    $0.7456$ &      $\pm$ &    $0.0043$ &     $0.58$ \\

  $^{13}$N &     $4682.0$ &    $\pm$ &       $4.9$ &     $0.10$ &   $0.5573$ &      $\pm$ &   $0.0013$ &     $0.23$ \\

  $^{15}$O &     $4402$ &      $\pm$ &       $11$ &     $0.25$ &  $-0.6281$ &      $\pm$ &   $0.0028$ &     $0.45$ \\

  $^{17}$F &     $2300.4$ &      $\pm$ &       $6.2$ &     $0.27$ &   $-1.2815$ &      $\pm$ &    $0.0035$ &     $0.27$ \\

 $^{19}$Ne &   $1718.4$ &      $\pm$ &      $3.2$ &     $0.19$ &    $1.5933$ &      $\pm$ &    $0.0030$ &     $0.19$ \\

 $^{21}$Na &     $4085$ &      $\pm$ &       $12$ &     $0.29$ &  $-0.7034$ &      $\pm$ &   $0.0032$ &     $0.45$ \\

 $^{23}$Mg &     $4725$ &      $\pm$ &       $17$ &     $0.36$ &   $0.5426$ &      $\pm$ &   $0.0044$ &     $0.81$ \\

 $^{25}$Al &   $3721.1$   &      $\pm$ &       $7.0$ &     $0.19$ &   $-0.7973$ &      $\pm$ &    $0.0027$ &     $0.34$ \\

 $^{27}$Si &     $4160$ &      $\pm$ &       $20$ &     $0.48$ &    $0.6812$ &      $\pm$ &    $0.0053$ &     $0.78$ \\

  $^{29}$P &     $4809$ &      $\pm$ &       $19$ &     $0.40$ &   $-0.5209$ &      $\pm$ &    $0.0048$ &     $0.92$ \\

  $^{31}$S &     $4828$ &      $\pm$ &       $33$ &     $0.68$ &    $0.5167$ &      $\pm$ &    $0.0084$ &     $1.63$ \\

 $^{33}$Cl &     $5618$ &      $\pm$ &       $13$ &     $0.23$ &    $0.3076$ &      $\pm$ &    $0.0042$ &     $1.37$ \\

 $^{35}$Ar &     $5688.6$ &    $\pm$ &      $7.2$ &     $0.13$ &  $-0.2841$ &      $\pm$ &   $0.0025$ &     $0.88$ \\

  $^{37}$K &     $4562$ &      $\pm$ &       $28$ &     $0.61$ &    $0.5874$ &      $\pm$ &    $0.0071$ &     $1.21$ \\

 $^{39}$Ca &     $4315$ &      $\pm$ &       $16$ &     $0.37$ &   $-0.6504$ &      $\pm$ &    $0.0041$ &     $0.63$ \\

 $^{41}$Sc &     $2849$ &      $\pm$ &       $11$ &     $0.39$ &  $-1.0561$ &      $\pm$ &   $0.0053$ &     $0.50$ \\

 $^{43}$Ti &     $3701$ &      $\pm$ &      $56$ &     $1.51$ &    $0.800$ &      $\pm$ &    $0.016$ &     $2.00$ \\

  $^{45}$V &     $4382$ &      $\pm$ &       $99$ &     $2.26$ &   $-0.621$ &      $\pm$ &    $0.025$ &     $4.03$ \\

  \hline
  \hline

\end{tabular}
\caption{  The $\ft^{\mbox{mirror}}$ values and Gamow-Teller/Fermi
mixing ratios, $\rho$, with their relative uncertainties.}

\label{Ft and rho}

\end{table}

\begin{figure}[!h]
\begin{center}
\includegraphics{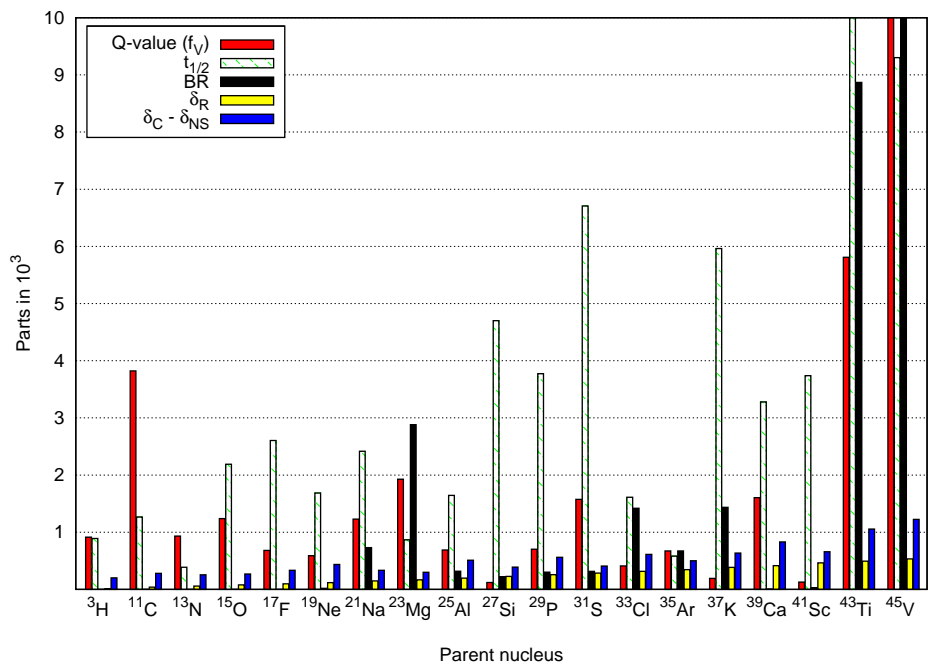}
\caption{   Histogram of the fractional uncertainties attributed
to each experimental and theoretical input factor that contributes
to the final $\mathcal{F} t^{mirror}$ values. }
\label{graph_relcontr}
\end{center}
\end{figure}

\newpage
\begin{sidewaystable}[!h]

\centering
\begin{tabular}{ | l   |  c |  r@{ }c@{ }l@{ $\quad$ } c@{ }   |     r@{ }c@{ }l@{$\quad$ } c@{ }   |      r@{ }c@{ }l@{$\quad$ } c@{ }  |  r@{ }c@{ }l@{ }  |  r@{ }c@{ }l@{ } |  }
\hline \hline
 &    & & &    & & &    &  &   &  &  &   &   &       &  & & &  \\
   parent &   spin &      \multicolumn{3}{c}{$a_{SM}$}  &  $\delta a$  &     \multicolumn{3}{c}{$A_{SM}$}  &  $\delta A$  &  \multicolumn{3}{c}{$B_{SM}$}  &  $\delta B$  & \multicolumn{3}{c}{$R^{FSI}$}  & \multicolumn{3}{c}{$N^{FSI}$} \\
   nucleus &  $J$ &     &   &        &                (\%)  &        &        &          &    (\%)    &       &    &            &      (\%)  &      &    &   & & &  \\
\hline
 &  &  & & &    & & &    &  &   &  &  &   &   &       &  & & &  \\

 $ ^{3}$H &        1/2 & $ -0.08593 $ &      $\pm$ & $ 0.00038 $ &   $ 0.44 $ & $ -0.09408 $ &      $\pm$ & $ 0.00046 $ &   $ 0.49 $ & $ 0.991849 $ &      $\pm$ & $ 0.000076 $ &   $ 0.01 $ & $ 0.005045 $ &      $\pm$ & $ 0.000025 $ & $ 0.09077 $ &      $\pm$ & $ 0.00044 $ \\

$ ^{11}$C  &        3/2 & $ 0.5236 $ &      $\pm$ & $ 0.0035 $ &   $ 0.67 $ & $ -0.59946 $ &      $\pm$ & $ 0.00016 $ &   $ 0.03 $ & $ -0.8853 $ &      $\pm$ & $ 0.0023 $ &   $ 0.26 $ & $ -0.008100 $ &      $\pm$ & $ 0.000006 $ & $ -0.20804 $ &      $\pm$ & $ 0.00012 $ \\

$ ^{13}$N  &        1/2 & $ 0.6840 $ &      $\pm$ & $ 0.0011 $ &   $ 0.16 $ & $ -0.333028 $ &      $\pm$ & $ 0.000040 $ &   $ 0.01 $ & $ -0.6490 $ &      $\pm$ & $ 0.0012 $ &   $ 0.18 $ & $ -0.004568 $ &      $\pm$ & $ 0.000001 $ & $ -0.099454 $ &      $\pm$ & $ 0.000022 $ \\

$ ^{15}$O  &        1/2 & $ 0.6228 $ &      $\pm$ & $ 0.0024 $ &   $ 0.39 $ & $ 0.7087 $ &      $\pm$ & $ 0.0022 $ &   $ 0.31 $ & $ 0.33148 $ &      $\pm$ & $ 0.00020 $ &   $ 0.06 $ & $ 0.008470 $ &      $\pm$ & $ 0.000027 $ & $ 0.16124 $ &      $\pm$ & $ 0.00051 $ \\

$ ^{17}$F  &        5/2 & $ 0.1713 $ &      $\pm$ & $ 0.0017 $ &   $ 0.99 $ & $ 0.99739 $ &      $\pm$ & $ 0.00018 $ &   $ 0.02 $ & $ 0.64222 $ &      $\pm$ & $ 0.00092 $ &   $ 0.14 $ & $ 0.013582 $ &      $\pm$ & $ 0.000003 $ & $ 0.226180 $ &      $\pm$ & $ 0.000049 $ \\

$ ^{19}$Ne  &        1/2 & $ 0.0435 $ &      $\pm$ & $ 0.0010 $ &   $ 2.30 $ & $ -0.04166 $ &      $\pm$ & $ 0.00095 $ &   $ 2.28 $ & $ -0.998186 $ &      $\pm$ & $ 0.000085 $ &   $ 0.01 $ & $ -0.000522 $ &      $\pm$ & $ 0.000012 $ & $ -0.00779 $ &      $\pm$ & $ 0.00018 $ \\

$ ^{21}$Na  &        3/2 & $ 0.5587 $ &      $\pm$ & $ 0.0027 $ &   $ 0.48 $ & $ 0.8614 $ &      $\pm$ & $ 0.0019 $ &   $ 0.22 $ & $ 0.59661 $ &      $\pm$ & $ 0.00032 $ &   $ 0.05 $ & $ 0.010731 $ &      $\pm$ & $ 0.000024 $ & $ 0.14457 $ &      $\pm$ & $ 0.00033 $ \\

$ ^{23}$Mg  &        3/2 & $ 0.6967 $ &      $\pm$ & $ 0.0044 $ &   $ 0.63 $ & $ -0.5584 $ &      $\pm$ & $ 0.0017 $ &   $ 0.30 $ & $ -0.7404 $ &      $\pm$ & $ 0.0040 $ &   $ 0.54 $ & $ -0.006529 $ &      $\pm$ & $ 0.000020 $ & $ -0.08023 $ &      $\pm$ & $ 0.00025 $ \\

$ ^{25}$Al  &        5/2 & $ 0.4818 $ &      $\pm$ & $ 0.0021 $ &   $ 0.44 $ & $ 0.9350 $ &      $\pm$ & $ 0.0011 $ &   $ 0.12 $ & $ 0.71289 $ &      $\pm$ & $ 0.00016 $ &   $ 0.02 $ & $ 0.011214 $ &      $\pm$ & $ 0.000013 $ & $ 0.12639 $ &      $\pm$ & $ 0.00014 $ \\

$ ^{27}$Si  &        5/2 & $ 0.5774 $ &      $\pm$ & $ 0.0053 $ &   $ 0.92 $ & $ -0.6959 $ &      $\pm$ & $ 0.0013 $ &   $ 0.19 $ & $ -0.8771 $ &      $\pm$ & $ 0.0032 $ &   $ 0.36 $ & $ -0.007899 $ &      $\pm$ & $ 0.000015 $ & $ -0.08230 $ &      $\pm$ & $ 0.00015 $ \\

$ ^{29}$P  &        1/2 & $ 0.7154 $ &      $\pm$ & $ 0.0048 $ &   $ 0.67 $ & $ 0.6154 $ &      $\pm$ & $ 0.0046 $ &   $ 0.75 $ & $ 0.33083 $ &      $\pm$ & $ 0.00044 $ &   $ 0.13 $ & $ 0.007298 $ &      $\pm$ & $ 0.000054 $ & $ 0.07059 $ &      $\pm$ & $ 0.00053 $ \\

$ ^{31}$S  &        1/2 & $ 0.7190 $ &      $\pm$ & $ 0.0084 $ &   $ 1.17 $ & $ -0.33043 $ &      $\pm$ & $ 0.00083 $ &   $ 0.25 $ & $ -0.6114 $ &      $\pm$ & $ 0.0080 $ &   $ 1.31 $ & $ -0.003804 $ &      $\pm$ & $ 0.000010 $ & $ -0.034356 $ &      $\pm$ & $ 0.000087 $ \\

$ ^{33}$Cl  &        3/2 & $ 0.8848 $ &      $\pm$ & $ 0.0029 $ &   $ 0.33 $ & $ -0.4007 $ &      $\pm$ & $ 0.0040 $ &   $ 1.00 $ & $ -0.4699 $ &      $\pm$ & $ 0.0057 $ &   $ 1.21 $ & $ -0.004739 $ &      $\pm$ & $ 0.000048 $ & $ -0.04010 $ &      $\pm$ & $ 0.00040 $ \\

$ ^{35}$Ar  &        3/2 & $ 0.9004 $ &      $\pm$ & $ 0.0016 $ &   $ 0.18 $ & $ 0.4371 $ &      $\pm$ & $ 0.0036 $ &   $ 0.82 $ & $ 0.3773 $ &      $\pm$ & $ 0.0026 $ &   $ 0.69 $ & $ 0.005102 $ &      $\pm$ & $ 0.000041 $ & $ 0.04063 $ &      $\pm$ & $ 0.00033 $ \\

$ ^{37}$K  &        3/2 & $ 0.6580 $ &      $\pm$ & $ 0.0061 $ &   $ 0.93 $ & $ -0.5739 $ &      $\pm$ & $ 0.0021 $ &   $ 0.37 $ & $ -0.7791 $ &      $\pm$ & $ 0.0058 $ &   $ 0.74 $ & $ -0.006863 $ &      $\pm$ & $ 0.000025 $ & $ -0.05158 $ &      $\pm$ & $ 0.00019 $ \\

$ ^{39}$Ca  &        3/2 & $ 0.6036 $ &      $\pm$ & $ 0.0041 $ &   $ 0.68 $ & $ 0.8270 $ &      $\pm$ & $ 0.0029 $ &   $ 0.35 $ & $ 0.58916 $ &      $\pm$ & $ 0.00076 $ &   $ 0.13 $ & $ 0.009766 $ &      $\pm$ & $ 0.000034 $ & $ 0.06950 $ &      $\pm$ & $ 0.00024 $ \\

$ ^{41}$Sc  &        7/2 & $ 0.2970 $ &      $\pm$ & $ 0.0033 $ &   $ 1.11 $ & $ 0.99777 $ &      $\pm$ & $ 0.00032 $ &   $ 0.03 $ & $ 0.76344 $ &      $\pm$ & $ 0.00080 $ &   $ 0.10 $ & $ 0.012480 $ &      $\pm$ & $ 0.000004 $ & $ 0.084287 $ &      $\pm$ & $ 0.000027 $ \\

$ ^{43}$Ti  &        7/2 &  $ 0.480 $ &      $\pm$ &  $ 0.016 $ &   $ 3.33 $ & $ -0.7737 $ &      $\pm$ & $ 0.0016 $ &   $ 0.21 $ & $ -0.9470 $ &      $\pm$ & $ 0.0057 $ &   $ 0.60 $ & $ -0.009563 $ &      $\pm$ & $ 0.000023 $ & $ -0.06147 $ &      $\pm$ & $ 0.00014 $ \\

$ ^{45}$V  &        7/2 &  $ 0.629 $ &      $\pm$ &  $ 0.021 $ &   $ 3.34 $ &  $ 0.852 $ &      $\pm$ &  $ 0.017 $ &   $ 2.00 $ &  $ 0.729 $ &      $\pm$ &  $ 0.010 $ &   $ 1.37 $ & $ 0.01060 $ &      $\pm$ & $ 0.00022 $ & $ 0.0650 $ &      $\pm$ & $ 0.0013 $ \\

\hline \hline
\end{tabular}
\caption{  Calculated standard model values for the $a, A, B, N$
and $R$ correlation coefficients for the $T = 1/2$ mirror $\beta$
transitions up to $^{45}$V, using the mixing ratios listed in
Table~V. The $D$~triple correlation is zero in the standard model.
The $\beta$ particle longitudinal polarization, $G$, is $-1$ for
$\beta^-$ decay and $+1$ for $\beta^+$ decay. The $N$ and $R$
correlations are non-zero due to final state interactions (FSI).
Note that the about 10\% accuracy to which the Eqs.~(32,33) used
to calculate $N^{FSI}$ and $R^{FSI}$ are valid \cite{vogel83} is
not included in the error bars.} \label{correlations}

\end{sidewaystable}



\normalsize


\newpage
\begin{longtable}[!h]{p{2.5cm}p{5cm}p{5cm}p{3cm}}

\caption{  References to data used in the calculation of the half-lives, $t_{1/2}$, of the $T = 1/2$ mirror nuclei. \label{usedref_half-life}}\\

\hline \hline \vspace{-0.2cm}
 &    &   &  \\
\vspace{-0.1cm} Code &    Authors &  Reference & Measured nuclei \\
 &    &   &  \\

\hline

           &            &            &            \\
\endfirsthead
\caption[]{   Continued}\\
\hline \hline \vspace{-0.2cm}
 &    &   &  \\
\vspace{-0.1cm} Code &    Authors &  Reference & Measured nuclei \\
 &    &   &  \\

\hline

           &            &            &            \\
\endhead
  &            &            &            \\
\hline
\endfoot
  &            &            &            \\
\hline \hline
\endlastfoot

    [Ak04] & Y.A. Akulov  \emph{et al.} & Phys. Lett. B \textbf{600}, 41 (2004) &    $^{3}$H \\

    [Ak88] & Y.A. Akulov  \emph{et al.} & Pis'ma Zh. Tekh. Fiz. \textbf{14}, 940-942 (1988). English translation: Sov. Tech. Phys. Lett. \textbf{14}, 416 (1988) &    $^{3}$H \\

    [Al67] & A.M. Aldridge \emph{et al.} & Nucl.Phys. A \textbf{98}, 323(1967) &  $^{43}$Ti \\

    [Al72] & D.E. Alburger, D.H. Wilkinson & Phys.Rev. C \textbf{6}, 2019 (1972) &   $^{17}$F \\

    [Al73] & D.E. Alburger, D.H. Wilkinson & Phys.Rev. C \textbf{8}, 657 (1973) & $^{39}$Ca, $^{41}$Sc \\

    [Al74] & D.E. Alburger & Phys.Rev. C \textbf{9}, 991 (1974) & $^{21}$Na, $^{23}$Mg, $^{31}$S \\

    [Al77] & D.E. Alburger & Phys.Rev. C \textbf{16}, 889 (1977) &   $^{17}$F \\

    [Ar58] & S.E. Arnell \emph{et al.} & Nucl.Phys. \textbf{6}, 196 (1958) & $^{11}$C, $^{13}$N, $^{21}$Na \\

    [Ar84] & Y. Arai \emph{et al.} & Nucl.Phys. A \textbf{420}, 193 (1984) &  $^{59}$Zn \\

    [Aw69] & M. Awschalom \emph{et al.} & Nucl.Instr.Methods \textbf{75}, 93 (1969) &   $^{11}$C \\

    [Ay84] & J. \"{A}yst\"{o} \emph{et al.} & Phys. Lett. B \textbf{138}, 369-372 (1984) & $^{51}$Fe, $^{55}$Ni \\

    [Az74] & G. Azuelos \emph{et al.} & Nucl. Instrum. Methods  \textbf{117}, 233 (1974) &  $^{23}$Mg \\

    [Az75] & G. Azuelos, J.E. Kitching & Phys.Rev. C  \textbf{12}, 563 (1975) & $^{11}$C, $^{19}$Ne, $^{21}$Na, $^{23}$Mg, $^{25}$Al, $^{27}$Si, $^{29}$P \\

    [Az77] & G. Azuelos \emph{et al.} & Phys.Rev. C  \textbf{15}, 1847 (1977) & $^{13}$N, $^{15}$O, $^{17}$F, $^{23}$Mg, $^{31}$S, $^{33}$Cl, $^{35}$Ar, $^{37}$K, $^{39}$Ca \\

    [Ba55] & S. Bashkin \emph{et al.} & Phys.Rev.  \textbf{99}, 107 (1955) &   $^{11}$C \\

    [Ba77] & P.H. Barker \emph{et al.} & Nucl.Phys. A \textbf{275}, 37 (1977) &  $^{27}$Si \\

    [Ba94] & P. Baumann \emph{et al.} & Phys.Rev. C  \textbf{50}, 1180 (1994) &  $^{67}$Se \\

    [Be75] & H. Behrens \emph{et al.} & Nucl.Phys. A  \textbf{246}, 317 (1975) &   $^{11}$C \\

    [Bl68] & J.L. Black, J. Mahieux & Nucl.Instr.Methods  \textbf{58}, 93 (1968) &  $^{27}$Si \\

    [Bl95] & B. Blank \emph{et al.} & Phys. Lett. B  \textbf{364}, 8 (1995) & $^{67}$Se, $^{71}$Kr \\

    [Bl99] &   B. Blank & J. Phys. G  \textbf{25}, 629 (1999) &  $^{79}$Zr \\

    [Bo65] & M. Bormann \emph{et al.} & Nucl.Phys.  \textbf{63}, 438 (1965) &   $^{13}$N \\

    [Bu85] & T.W. Burrows \emph{et al.} & Phys. Rev. C  \textbf{31}, 1490 (1985) &  $^{47}$Cr \\

    [Bu91] & B. Budick  \emph{et al.} & Phys. Rev. Lett.  \textbf{67}, 2630-2633 (1991) &    $^{3}$H \\

    [Cl58] & J.E. Cline, P.R. Chagnon & Bull. Am. Phys. Soc. 3, No.3, 206, RA5 (1958) & $^{31}$S, $^{39}$Ca \\

    [Di51] & J.M. Dickson, T.C. Randle & Proc. Phys. Soc. (London) \textbf{64}A, 902 (1951) &   $^{11}$C \\

    [Ea62] & L.G. Earwaker \emph{et al.} & Nature \textbf{195}, 271 (1962) &  $^{19}$Ne \\

    [Eb65] & T.G. Ebrey, P.R. Gray & Nucl.Phys. \textbf{61}, 479 (1965) &   $^{13}$N \\

    [Ed77] & M.D. Edmiston \emph{et al.} & Nucl. Instrum. Methods \textbf{141}, 315 (1977) &  $^{47}$Cr \\

    [Ew81] & G.T. Ewan \emph{et al.} & Nucl.Phys. A \textbf{352}, 13 (1981) &  $^{71}$Kr \\

    [Ge76] & H. Genz \emph{et al.} & Nucl.Instrum.Methods \textbf{134}, 309 (1976) &  $^{27}$Si \\

    [Go68] & J.D.Goss \emph{et al.} & Nucl.Phys. A \textbf{115}, 113 (1968) & $^{19}$Ne, $^{23}$Mg, $^{27}$Si \\

    [Gr71] & D. Grober, W. Gruhle & BMBW-FBK-71-09, p.90 (1971) &  $^{31}$Si \\

    [Ha52] & R.N.H. Haslam \emph{et al.} & Can.J.Phys. \textbf{30}, 257 (1952) &   $^{31}$S \\

    [Ha80] & J.C. Hardy \emph{et al.} & Phys. Lett. B \textbf{91}, 207 (1980) &  $^{49}$Mn \\

    [Ha87] & H. Hama \emph{et al.} & Proc. 5th Int.Conf.Nuclei Far from Stability, Rosseau Lake, Canada 1987, Ed., I.S.Towner, p.650 (1988) & $^{45}$V, $^{47}$Cr, $^{49}$Mn, $^{51}$Fe, $^{53}$Co, $^{55}$Ni \\

    [Ho77] & P. Hornshoj \emph{et al.} & Nucl. Phys. A \textbf{288}, 429 (1977) &  $^{55}$Ni \\

    [Ho82] & P. Hornshoj \emph{et al.} & Phys.Lett. B \textbf{116}, 4 (1982) &   $^{45}$V \\

    [Ho87] & J. Honkanen \emph{et al.} & Nucl.Phys. A \textbf{471}, 489 (1987) &  $^{43}$Ti \\

    [Ho89] & J. Honkanen \emph{et al.} & Nucl.Phys. A \textbf{496}, 462 (1989) &  $^{53}$Co \\

    [Hu03] & J. Huikari \emph{et al.} & Eur. Phys. J. A\textbf{16}, 359 (2003) &  $^{75}$Sr \\

    [Hu54] & S.E. Hunt \emph{et al.} & Phys. Rev. \textbf{95}, 611A (1954)  &   $^{31}$S \\

    [Ia06] & V.E. Iacob \emph{et al.} & Phys.Rev. C \textbf{74}, 055502 (2006) &  $^{35}$Ar \\

    [Ja60] & J. Janecke & Z.Naturforsch. \textbf{15a}, 593 (1960) & $^{13}$N, $^{15}$O, $^{29}$P, $^{31}$S, $^{33}$Cl, $^{35}$Ar, $^{41}$Sc, $^{43}$Ti \\

    [Ja61] & J. Janecke, H. Jung & Z. Phys. \textbf{165}, 94 (1961) &  $^{43}$Ti \\

    [Je50] & G.H. Jenks \emph{et al.} & Phys. Rev. \textbf{80}, 990-995 (1950) &    $^{3}$H \\

    [Jo51] & W.M. Jones & Phys. Rev. \textbf{83}, 537-539 (1950) &    $^{3}$H \\

    [Jo55] & W.M. Jones & Phys. Rev. \textbf{100}, 124-125 (1955) &    $^{3}$H \\

    [Jo67] & P.M.S. Jones & J. Nucl. Mater. \textbf{21}, 239-240 (1967) &    $^{3}$H \\

    [Ju71] & F. Jundt \emph{et al.} & Nucl.Phys. A \textbf{170}, 12 (1971) &  $^{25}$Al \\

    [Ka64] & R.W. Kavanagh, D.R. Goosman & Phys. Lett. \textbf{12}, 229 (1964); Erratum Phys.Lett. \textbf{13}, 358 (1964) &   $^{37}$K \\

   [Ka64a] & R.W. Kavanagh \emph{et al.} & Can. J. Phys. \textbf{42}, 1429 (1964) &   $^{11}$C \\

    [Ka68] & J.A. Kadlecek & Bull. Am. Phys. Soc. \textbf{13}, 676, HF15 (1968) &  $^{39}$Ca \\

    [Ki01] & K. Kienle \emph{et al.} & Prog. Part. Nucl. Phys. \textbf{46} (2001)73 & $^{77}$Y, $^{79}$Zr, $^{83}$Mo \\

    [Ki59] & O.C. Kistner, B.M. Rustad & Phys.Rev. \textbf{114}, 1329(1959) &   $^{15}$O \\

    [Ki60] & J.D. King \emph{et al.} & Can. J. Phys. \textbf{38}, 231 (1961) &   $^{13}$N \\

    [Kl54] & R.M. Kline, D.J. Zaffarano & Phys.Rev. 96, 1620 (1954) &  $^{39}$Ca \\

    [Ko73] & S. Kochan \emph{et al.}. & Nucl. Phys. A \textbf{204}, 185 (1973) &  $^{53}$Co \\

    [Ku53] & D. N. Kundu \emph{et al.} & Phys. Rev. \textbf{89}, 1200 (1953) &   $^{11}$C \\

    [Li60] & K.H. Lindenberger, J.A. Scheer & Z.Physik \textbf{158}, 111 (1960) & $^{31}$S, $^{39}$Ca \\

    [Lo02] & M. Lopez-Jimenez, B. Blank \emph{et al.} & Phys. Rev. C \textbf{66}, 025803 (2002) & $^{53}$Co, $^{55}$Ni, $^{57}$Cu, $^{59}$Zn, $^{61}$Ga, $^{63}$Ge, $^{65}$As, $^{67}$Se, $^{71}$Kr \\

    [Lu00] & L.L. Lucas and M.P. Unterweger & J. Res. Natl. Inst. Stand. Technol. \textbf{105}, 541 (2000) &    $^{3}$H \\

    [Ma06] & D. MacMahon & Appl. Rad. Isot. \textbf{64}, 1417-1419 (2006) &    $^{3}$H \\

    [Me66] & J.S. Merritt and J.G.V. Taylor & Report AECL-2510, Atomic Energy of Canada Limited, Chalk River Laboratory, Chalk River, Ontario (1966), p28 &    $^{3}$H \\

    [Mi58] & M.V. Mihailovic, B. Povh & Nuclear Phys. \textbf{7}, 296 (1958) & $^{27}$Si, $^{39}$Ca \\

    [Mo95] & D.J. Morrissey & Nucl. Phys. A \textbf{588}, c203 (1995) &  $^{65}$As \\

    [Mu58] & T. Muller \emph{et al.} & Physica \textbf{24}, 577 (1958) & $^{25}$Al, $^{33}$Cl \\

    [Ne63] & J.W. Nelson \emph{et al.} & Phys.Rev. \textbf{129}, 1723 (1963) & $^{15}$O, $^{31}$S, $^{35}$Ar \\

    [No47] &  A. Novick & Phys. Rev. \textbf{72}, 972 (1947) &    $^{3}$H \\

    [Oi97] & M.Oinonen \emph{et al.} & Phys.Rev. C \textbf{56}, 745 (1997) &  $^{71}$Kr \\

    [Ol87] & B.M. Oliver  \emph{et al.} & Appl. Radiat. Isot. \textbf{38}, 959-965 (1987) &    $^{3}$H \\

    [Pe57] & J.R. Penning, F.H. Schmidt & Phys.Rev. \textbf{105}, 647(1957) & $^{15}$O, $^{19}$Ne \\

    [Pi85] & L. E. Piilonen & PhD thesis, Princeton University &  $^{19}$Ne \\

    [Pl62] & H.S. Plendl \emph{et al.} & Conf. Low Energy Nuclear Phys. Harwell (september 1962): AERE-R-4131, 22 (1962) abstr.7a8 &  $^{43}$Ti \\

    [Po59] & M.M. Povov \emph{et al.} & Atomnaya Energiya 4, 196-298 (1958). English translations: Soviet J. At. Energy 4, 393-396 (1958) and J. Nucl. Energy 9, 190-193 (1959) &    $^{3}$H \\

    [Pr57] & I.D. Prokoshkin, A.A. Tiapkin & Zhur. Eksptl.I Teoret.Fiz. \textbf{32},117 (1957); Soviet Phys. JETP \textbf{5}, 148 (1957) &   $^{11}$C \\

    [Re99] & I. Reusen \emph{et al.} & Phys. Rev. C \textbf{59}, 2416 (1999) &  $^{55}$Ni \\

    [Ri68] & A.I.M. Ritchie & Nucl.Instr.Methods \textbf{64}, 181 (1968)  &   $^{13}$N \\

    [Ru77] & C.R. Rudy and K.C. Jordan & Progress Report MLM-2458, U.S. Department of Energy, Mound Laboratory, Miamisburg, Ohio, December 1977, pp. 2-10 &    $^{3}$H \\

    [Sc48] & A.D. Schelberg \emph{et al.} & Rev. Sci. Instr. \textbf{19}, 458 (1948) &  $^{43}$Ti \\

    [Sc58] & F. Schweizer & Phys.Rev. \textbf{110}, 1414 (1958) &   $^{37}$K \\

    [Sc70] & P.J. Scanlon, D. Crabtree & Can.J.Phys. \textbf{48}, 1578 (1970) & $^{29}$P, $^{33}$Cl \\

    [Se96] & D.R. Semon \emph{et al.} & Phys.Rev. C \textbf{53}, 96 (1996) &  $^{57}$Cu \\

    [Sh89] & T. Shinozuka \emph{et al.} & Proc. XXIII Yamada Conference on Nuclear Weak Process and Nuclear Structure, Osaka 1989, Eds. M.Morita, H.Ejiri, H.Ohtsubo, and T.Sato (World Scientific, Singapore, 1989), p. 108 &  $^{57}$Cu \\

    [Sh93] & B.M. Sherrill \emph{et al.} & Proc. 6th Int. Conf. On Nuclei Far from Stability + 9th Conf. On Atomic Masses and Fundamental Constants, Germany 1992, R. Neugard, A.Wohr eds. , 891 (1993) &  $^{63}$Ge \\

    [Si87] & J.J. Simpson & Phys. Rev. C \textbf{35}, 752-754 (1987) &    $^{3}$H \\

    [Sm41] & J.H.C. Smith, D.B. Cowie & J. Appl. Phys. \textbf{12}, 78 (1941) &   $^{11}$C \\

    [Su62] & D.C. Sutton & Thesis, Princeton University (1962) &  $^{27}$Si \\

    [Ta73] & I.Tanihata \emph{et al.} & J.Phys.Soc.Jap. \textbf{34}, 848 (1973) & $^{25}$Al, $^{29}$P, $^{33}$Cl, $^{41}$Sc \\

    [Un00] & M.P. Unterweger and L. L. Lucas & Appl. Radiat. Isot. \textbf{52}, 527-531 (2000) &    $^{3}$H \\

    [Va63] & S.S. Vasilev, L.Y. Shavtvalov & Zhur. Eksperim. I Teor. Fiz. \textbf{45}, 1385 (1963), Soviet Phys. JETP \textbf{18}, 995 (1964) &  $^{43}$Ti \\

    [Va69] & S.S. Vasilev \emph{et al.} & Vestn.Mosk.Univ., Fiz., Astron. No.5, 3 (1969) &  $^{43}$Ti \\

    [Wa60] & R. Wallace, J.A. Welch,Jr. & Phys.Rev. \textbf{117}, 1297 (1960) &   $^{31}$S \\

    [We02] & L. Weissman \emph{et al.} & Phys.Rev. C \textbf{65}, art. No. 044321 &  $^{61}$Ga \\

    [Wi69] & G.L. Wick \emph{et al.} & Nucl.Phys. A \textbf{138}, 209 (1969) &  $^{35}$Ar \\

    [Wi74] & D.H. Wilkinson, D.E. Alburger & Phys.Rev. C \textbf{10}, 1993 (1974) &  $^{19}$Ne \\

    [Wi80] & H.S. Wilson \emph{et al.} & Phys.Rev. C \textbf{22}, 1696 (1980) & $^{29}$P, $^{31}$S \\

    [Wi93] & J.A. Winger \emph{et al.} & Phys.Lett. B \textbf{299}, 214 (1993), Phys. Rev. C \textbf{48}, 3097 (1993) & $^{61}$Ga, $^{63}$Ge, $^{65}$As \\

    [Wo02] & D.H. Woods \emph{et al.} & App. Rad. Isot. \textbf{56}, 327 (2002) &   $^{11}$C \\

    [Wo69] & V.K. Wohlleben, E. Schuster & Radiochim.Acta \textbf{12}, 75 (1969) &   $^{17}$F \\

    [Yo65] & D.H. Youngblood \emph{et al.} & Nucl.Phys. \textbf{65}, 602(1965) &  $^{41}$Sc \\

\end{longtable}

\newpage
\begin{longtable}[!h]{p{2.5cm}p{5cm}p{5cm}p{3cm}}
\caption{   References to data neglected in the calculation of the
half-lives, $t_{1/2}$, of the mirror nuclei, with
the reason for their rejection.\label{unusedref_half-life}}\\
\hline \hline \vspace{-0.2cm}
 &    &   &  \\
\vspace{-0.1cm} Code &    Authors &  Reference & Measured nuclei \\
 &    &   &  \\

\hline

           &            &            &            \\
\endfirsthead
\caption[]{Continued}\\
\hline \hline \vspace{-0.2cm}
 &    &   &  \\
\vspace{-0.1cm} Code &    Authors &  Reference & Measured nuclei \\
 &    &   &  \\

\hline

           &            &            &            \\
\endhead
  &            &            &            \\
\hline
\endfoot
  &            &            &            \\
\hline \hline
\endlastfoot

\multicolumn{4}{l}{\bf Error bar 10 times higher than most precise measurement}            \\

    [Al57] & W.P. Alford, D.R. Hamilton & Phys. Rev. \textbf{105}, 673 (1957) &  $^{19}$Ne \\

    [Al59] & J.S. Allen \emph{et al.} & Phys. Rev. \textbf{116}, 134 (1959) & $^{19}$Ne, $^{35}$Ar \\

    [Ar58] & S.E. Arnell \emph{et al.} & Nucl. Phys. \textbf{6}, 196 (1958) & $^{25}$Al  \\

    [Ar81] & Y. Arai \emph{et al.} & Phys. Lett. B \textbf{104},186 (1981) &  $^{59}$Zn \\

    [Ba64] & J.E.E. Baglin, B.M. Spicer & Nucl. Phys. \textbf{54}, 549 (1964) &  $^{39}$Ca \\

    [Ba70] & T.T. Bardin \emph{et al.} & Phys. Rev. C \textbf{2}, 2283 (1970) &  $^{33}$Cl \\

    [Bl51] & J.P. Blaser \emph{et al.} & Helv. Phys. Acta \textbf{24}, 441 (1951) &  $^{27}$Si \\

    [Bl95] & B. Blank \emph{et al.} & Phys. Lett. B  \textbf{364}, 8 (1995) &  $^{75}$Sr \\

    [Bo53] & F.I. Boley & Iowa State Coll. J. Sci. \textbf{27},129 (1953) & $^{21}$Na, $^{23}$Mg, $^{27}$Si, $^{31}$S, $^{33}$Cl, $^{37}$K, $^{39}$Ca \\

    [Br53] & R. Braams, C.L. Smith & Phys. Rev. \textbf{90}, 995 (1953) &  $^{39}$Ca \\

    [Bu65] & I.F. Bubb \emph{et al.} & Nucl. Phys.\textbf{65}, 655 (1965) &  $^{27}$Si \\

    [Ch53] & J.L.W. Churchill \emph{et al.} & Nature \textbf{172}, 460 (1953) &  $^{25}$Al \\

    [Cl58] & J.E. Cline, P.R. Chagnon & Bull. Am. Phys. Soc. 3, No.3, 206, RA5 (1958) &  $^{27}$Si \\

    [Cr40] & E.C. Creutz \emph{et al.} & Phys. Rev. \textbf{57}, 567 (1940) & $^{21}$Na, $^{27}$Si \\

    [Cr62] & J.G. Cramer Jr., C.M. Class & Nucl. Phys. \textbf{34}, 580 (1962) &  $^{41}$Sc \\

    [Cs63] & J. Csikai, G. Peto & Phys.Letters \textbf{4}, 252 (1963) &   $^{15}$O \\

    [El41] & D.R. Elliott, L.D. King & Phys. Rev. \textbf{60},  489 (1941) & $^{27}$Si, $^{35}$Ar, $^{41}$Sc \\

    [Es72] & M.A. Eswaran \emph{et al.} & Phys. Rev. C \textbf{5}, 1270 (1972) &  $^{33}$Cl \\

    [Fr69] & J.M. Freeman \emph{et al.} & Phys. Lett. B\textbf{5}, 296 (1969) &  $^{35}$Ar \\

    [Ge71] & J.S. Geiger, B.W. Hooton & Can. J. Phys. \textbf{49}, 663 (1971) &  $^{35}$Ar \\

    [Go64] & S. Gorodetzky \emph{et al.} & Compt. Rend. Congr. Intern. Phys. Nucl., Paris, P.Gugenberger, Ed., Centre National de la Recherche Scientifique, Paris, Vol.II, p.408 (1964) &  $^{27}$Si \\

    [Gr71] & D. Grober, W. Gruhle & BMBW-FBK-71-09, p.90 (1971) & $^{25}$Al, $^{29}$P \\

    [Ho73] & K.R. Hogstrom \emph{et al.} & Nucl. Phys. A \textbf{215}, 598 (1973) &   $^{11}$C \\

    [Ho77] & P. Hornshoj \emph{et al.} & Nucl. Phys. A \textbf{288}, 429 (1977) &  $^{47}$Cr \\

    [Ho81] & J. Honkanen \emph{et al.} & Nucl. Phys. A \textbf{366}, 109 (1981) &  $^{59}$Zn \\

    [Hu41] &   P. Huber & Helv. Phys. Acta \textbf{14}, 163 (1941) &   $^{31}$S \\

    [Hu43] & O. Huber \emph{et al.} & Helv. Phys. Acta \textbf{16}, 33 (1943) & $^{23}$Mg (1,06 of 1,08) \\

    [Hu44] & O. Huber \emph{et al.} & Helv. Phys. Acta \textbf{17}, 195 (1944) &  $^{27}$Si \\

    [Hu54] & S.E. Hunt \emph{et al.} & Phys. Rev. \textbf{95}, 611A (1954)  & $^{25}$Al, $^{27}$Si \\

    [Ja60] & J. Janecke & Z. Naturforsch. \textbf{15}A, 593 (1960) &  $^{25}$Al \\

    [Ki01] & K. Kienle \emph{et al.} & Prog. Part. Nucl. Phys. \textbf{46}, 73 (2001) &  $^{75}$Sr, $^{79}$Zr \\

    [Ki56] & O.C. Kistner \emph{et al.} & Phys. Rev. \textbf{104}, 154 (1956) &  $^{35}$Ar \\

    [La48] & R.V. Langmuir & Phys. Rev. \textbf{74}, 1559A (1948) &   $^{37}$K \\

    [Lo02] & M.J. Lopez-Jimenez \emph{et al.} & Phys. Rev. C \textbf{66}, 025803 (2002) & $^{53}$Co, $^{55}$Ni, $^{57}$Cu, $^{59}$Zn, $^{61}$Ga, $^{63}$Ge, $^{65}$As, $^{67}$Se, $^{71}$Kr \\

    [Mc49] & J. McElhinney \emph{et al.} & Phys. Rev. \textbf{75}, 542 (1949) &   $^{31}$S \\

    [Mo71] & C.E. Moss \emph{et al.} & Nucl. Phys. A \textbf{170}, 111 (1971) & $^{25}$Al, $^{37}$K \\

    [Na54] & M.E. Nahmias & J. Phys. Radium \textbf{15}, 677 (1954) &  $^{19}$Ne \\

    [Ne63] & J.W. Nelson \emph{et al.} & Phys. Rev. \textbf{129}, 1723 (1963) & $^{15}$O, $^{31}$S, $^{35}$Ar \\

    [Pa65] & J.R. Patterson \emph{et al.} & Proc. Phys. Soc. (London) \textbf{86}, 1297 (1965) &   $^{11}$C \\

    [Ph53] & P. Phipps, D.J. Zaffarano & ISC-443 (1953) &  $^{21}$Na \\

    [Ri55] & C.S. Ring Jr., D.J. Zaffarano & ISC-648 (1955) & $^{39}$Ca,  \\

    [Sc52] & G. Schrank, J.R. Richardson & Phys. Rev. \textbf{86}, 248-248 (1952) & $^{19}$Ne, $^{21}$Na \\

    [Sh84] & T. Shinozuka \emph{et al.} & Phys. Rev. C \textbf{30}, 2111 (1984) &  $^{57}$Cu \\

    [Si44] & K. Siegbahn & Arkiv. Mat. Astron. Fysik \textbf{30}A, no. 20 (1944) &   $^{11}$C \\

    [Si73] &   J. Singh & Proc.Nucl.Phys.and Solid State Phys.Symp., Chandigarh, Vol.15B, p.1 (1973) & $^{11}$C, $^{13}$N \\

    [So41] & A.K. Solomon & Phys. Rev. \textbf{60},  279 (1941) &   $^{11}$C \\

    [Su53] & R.G. Summers-Gill \emph{et al.} & Can. J. Phys. \textbf{31}, 70 (1953) &  $^{27}$Si \\

    [Su58] & C.R. Sun, B.T. Wright & Phys. Rev. \textbf{109}, 109 (1958) &   $^{37}$K \\

    [Ty54] & H. Tyren, P.A. Tove & Phys. Rev. \textbf{96}, 773 (1954) &  $^{43}$Ti \\

    [Va60] & S.S. Vasilev, L.Y. Shavtvalov & Zhur. Eksptl. I teoret. Fiz. \textbf{39}, 1221 (1960), Soviet Phys. JETP \textbf{12}, 851 (1961) &  $^{27}$Si \\

    [Va62] & S.S. Vasilev, L.Y. Shavtvalov & Izvest. Akad. Anuk SSSR, Ser. Fiz. \textbf{26}, 1495 (1962); Columbia Tech. Transl. \textbf{26}, 1521 (1963) & $^{17}$F, $^{33}$Cl \\

    [Wa60] & R. Wallace, J.A. Welch,Jr. & Phys. Rev. \textbf{117}, 1297 (1960) & $^{21}$Na, $^{23}$Mg, $^{25}$Al, $^{27}$Si, $^{29}$P, $^{33}$Cl, $^{35}$Ar, $^{37}$K, $^{39}$Ca, $^{41}$Sc \\

    [Wh39] & M.G. White \emph{et al.} & Phys. Rev. \textbf{56}, 512-518 (1939) & $^{19}$Ne, $^{23}$Mg \\

    [Wh41] & M.G. White \emph{et al.} & Phys. Rev. \textbf{59}, 63-68 (1941) & $^{29}$P, $^{31}$S, $^{33}$Cl, $^{35}$Ar \\

           &            &            &            \\

\multicolumn{4}{l}{\bf No error bar quoted, and/or no definite value, merely a limit.}             \\

    [Ho40] &  J.B. Hoag & Phys. Rev. \textbf{57}, 937 (1940) &  $^{33}$Cl \\

    [Ma52] & W.M. Martin, S.W. Breckon & Can. J. Physics \textbf{30}, 64 (1952) &  $^{39}$Ca \\

    [Bl98] &   B. Blank & J. Phys. G \textbf{24}, 1385 (1998) &   $^{77}$Y \\

    [Ki01] & K. Kienle \emph{et al.} & Prog. Part. Nucl. Phys. \textbf{46}, 73 (2001) &  $^{81}$Nb \\

           &            &            &            \\

\multicolumn{4}{l}{\bf Updated in [Re95]}     \\

    [Re95] & I. Reusen \emph{et al.} & Proc. Intern. Conf. On exotic Nuclei and Atomic Masses, Arles 1995,757 &  $^{55}$Ni \\

    [Ve94] & L. Vermeeren \emph{et al.} & Phys. Rev. Lett. \textbf{73}, 1935 (1994) &  $^{55}$Ni \\

           &            &            &            \\


  \multicolumn{4}{l}{\bf  Pre-1958 data that are systematically higher than later and equally precise results.}         \\

    [Ch53] & J.L.W. Churchill \emph{et al.} & Nature \textbf{172}, 460 (1953) &   $^{13}$N \\

    [Da57] & H. Daniel, U. Schmidt-Rohr & Z. Naturforsch. \textbf{12}A, 750 (1957) &   $^{13}$N \\

    [De57] & A.S. Deineko \emph{et al.} & Zhur. Eksptl.I Teoret.Fiz. \textbf{32}, 251 (1957); Soviet Phys. JETP \textbf{5}, 201 (1957) &   $^{13}$N \\

    [Ho50] & W.F. Hornyak \emph{et al.} & Rev. Mod. Phys. \textbf{22}, 291 (1950), Phys. Rev. \textbf{77},160 (1950) &   $^{13}$N \\

    [No57] & E. Norbeck Jr., C.S. Littlejohn & Phys. Rev. \textbf{108}, 754 (1957) &   $^{13}$N \\

    [Si45] & K. Siegbahn, Slaetis & Arkiv. Mat. Astron. Fysik \textbf{32}A, no. 9 (1945) &   $^{13}$N \\

    [Wa39] &       Ward & Proc. Camb. Phil. Soc. \textbf{35}, 523 (1939) &   $^{13}$N \\

    [Wi55] & D.H. Wilkinson & Phys. Rev. \textbf{100}, 32 (1955) &   $^{13}$N \\

    &            &            &            \\

\multicolumn{4}{l}{\bf  Strongly deviating values, possible contamination.}         \\

    [Ba55] & S. Bashkin \emph{et al.} & Phys. Rev.  \textbf{99}, 107 (1955) &   $^{15}$O \\

    [Br50] & H. Brown and V. Perez-Mendez  & Phys. Rev. \textbf{78}, 649 (1950) &   $^{15}$O \\

    [Ki57] & O.C. Kistner \emph{et al.} & Phys. Rev. \textbf{105}, 1339 (1957) &   $^{15}$O \\

    [Kl54] & R.M. Kline, D.J. Zaffarano & Phys. Rev. \textbf{96}, 1620 (1954) &   $^{15}$O \\

    &            &            &            \\

\multicolumn{4}{l}{\bf  Pre-1969 data that are systematically higher, possibility of $^{15}$O contamination.}         \\

    [Ar58] & S.E. Arnell \emph{et al.} & Nucl. Phys. \textbf{6}, 196 (1958) &  $^{17}$F  \\

    [Br49] & H. Brown and V. Perez-Mendez & Phys. Rev. \textbf{75},1286A (1949) &   $^{17}$F \\

    [Ja60] & J. Janecke & Z.Naturforsch. \textbf{15}A, 593 (1960) &   $^{17}$F \\

    [Ko54] & L. Koester & Z. Naturforsch. \textbf{9}A, 104 (1954) &   $^{17}$F \\

    [La51] & R.A. Laubenstein \emph{et al.} & Phys. Rev. \textbf{84}, 12 (1951)  &   $^{17}$F \\

    [Wo54] &    C. Wong & Phys. Rev. \textbf{95}, 765-766 (1954) &   $^{17}$F \\

    &            &            &            \\

\multicolumn{4}{l}{\bf  Strongly deviating measurements.}         \\

    [Ja60] & J. Janecke & Z. Naturforsch. \textbf{15}A, 593 (1960) &  $^{19}$Ne \\

    [Va64] & S.S. Vasilev \emph{et al.} & Zhur. Eksperim. I Teor. Fiz. \textbf{47}, 1164 (1964), Soviet Phys. JETP \textbf{20}, 783 (1965) &  $^{19}$Ne \\

    [Wa60] & R. Wallace, J.A. Welch,Jr. & Phys. Rev. \textbf{117}, 1297 (1960) &  $^{19}$Ne \\

           &            &            &            \\

    [Mi58] & M.V. Mihailovic, B. Povh & Nuclear Phys. \textbf{7}, 296 (1958) &  $^{23}$Mg \\

           &            &            &            \\

    [Ro55] & H. Roderick \emph{et al.} & Phys. Rev. \textbf{97}, 97-101 (1955) &   $^{29}$P \\

           &            &            &            \\

    [El41] & D.R. Elliott, L.D. King & Phys. Rev. \textbf{60},  489 (1941) &   $^{31}$S \\

    [Mi58] & M.V. Mihailovic, B. Povh & Nucl. Phys. \textbf{7}, 296 (1958) &   $^{31}$S \\

           &            &            &            \\

    [El41] & D.R. Elliott, L.D. King & Phys. Rev. \textbf{60},  489 (1941) &  $^{41}$Sc \\

    [Wa60] & R. Wallace, J.A. Welch,Jr. & Phys. Rev. \textbf{117}, 1297 (1960) &  $^{41}$Sc \\

[Ho77] & P. Hornshoj \emph{et al.} & Nucl. Phys. A \textbf{288}, 429 (1977) &  $^{51}$Fe \\

\end{longtable}

\newpage
\begin{table}[!h]

\begin{tabular}{lp{5cm}  p{7cm}  p{3cm} }

\hline \hline
 & & & \\
 Code & Authors & Reference & Nucleus \\

 \hline

 & & & \\

    [Ac07] & N. Achouri and O. Naviliat-Cuncic & private communication & $^{21}$Na\\

    [Ad81] & E.G. Adelberger \emph{et al.} & Phys. Rev. C \textbf{24},313 (1981) &  $^{19}$Ne \\

    [Ad83] & E.G. Adelberger \emph{et al.} & Phys. Rev. C \textbf{27},2833 (1983) &  $^{19}$Ne \\

    [Ad84] & E.G. Adelberger \emph{et al.} & Nucl. Phys. A \textbf{417}, 269 (1984) &  $^{35}$Ar \\

    [Al74] & D.E. Alburger & Phys.Rev. C \textbf{9}, 991 (1974) & $^{21}$Na, $^{23}$Mg, $^{31}$S \\

    [Al76] & D.E. Alburger & Phys. Rev. C \textbf{13}, 2593 (1976) &  $^{19}$Ne \\

    [Ar64] & S.E. Arnell, E. Wernbom & Arkiv Fysik \textbf{25}, 389 (1964) & $^{17}$F, $^{21}$Na, $^{25}$Al \\

    [Ar84] & Y. Arai \emph{et al.} & Nucl.Phys. A \textbf{420}, 193 (1984) &  $^{59}$Zn \\

    [Ay84] & J. \"{A}yst\"{o} \emph{et al.} & Phys. Lett. B \textbf{138}, 369-372 (1984) &  $^{51}$Fe \\

    [Az77] & G. Azuelos \emph{et al.} & Phys.Rev. C \textbf{15}, 1847 (1977) & $^{21}$Na, $^{23}$Mg, $^{25}$Al, $^{27}$Si, $^{29}$P, $^{31}$S, $^{35}$Ar, $^{37}$K \\

    [Be71] & D. Berenyi \emph{et al.} & Nucl. Phys. A \textbf{178}, 76 (1971) &  $^{27}$Si \\

    [Bu85] & T.W. Burrows \emph{et al.} & Phys. Rev. C \textbf{31}, 1490 (1985) &  $^{47}$Cr \\

    [De71] & C. D\'{e}traz \emph{et al.} & Phys. Lett. B \textbf{34},128 (1971) & $^{27}$Si, $^{31}$S, $^{35}$Ar \\

    [Ia06] & V.E. Iacob \emph{et al.} & Phys.Rev. C \textbf{74} (2006) 015501 &  $^{21}$Na \\

    [Ge71] & J.S. Geiger, B.W. Hooton & Can. J. Phys. \textbf{49}, 663 (1971) &  $^{35}$Ar \\

   [Go68a] & S. Gorodetzky \emph{et al.} & Nucl. Phys. A \textbf{109}, 417 (1968) &  $^{23}$Mg \\

    [Ha80] & J.C. Hardy \emph{et al.} & Phys. Lett. B \textbf{91}, 207 (1980) &  $^{49}$Mn \\

    [Ha94] & E. Hagberg \emph{et al.} & Nucl. Phys. A \textbf{571}, 555 (1994) &  $^{39}$Ca \\

    [Ha97] & E. Hagberg \emph{et al.} & Phys. Rev. C \textbf{56}, 135 (1997) &   $^{37}$K \\

    [Ho50] & W.F. Hornyak \emph{et al.} & Rev. Mod. Phys. \textbf{22}, 291 (1950), Phys. Rev.\textbf{77},160 (1950) &   $^{11}$C \\

    [Ho82] & P. Hornshoj \emph{et al.} & Phys.Lett. B \textbf{116}, 4 (1982) &   $^{45}$V \\

    [Ho87] & J. Honkanen \emph{et al.} & Nucl.Phys. A \textbf{471}, 489 (1987) &  $^{43}$Ti \\

    [Ho89] & J. Honkanen \emph{et al.} & Nucl.Phys. A \textbf{496}, 462 (1989) & $^{49}$Mn, $^{51}$Fe, $^{53}$Co \\

    [Lo62] &  O. Lonsjo & Phys. Norvegica \textbf{1}, 41 (1962) &   $^{29}$P \\

    [Ma69] & L. Makela \emph{et al.} & Bull. Am. Phys. Soc. \textbf{14}, 550  (1969) &  $^{25}$Al \\

    [Ma74] & F.M. Mann, R.W. Kavanagh & Nucl. Phys. A \textbf{235}, 299 (1974) & $^{23}$Mg, $^{27}$Si \\

    [Ma76] & F.M. Mann \emph{et al.} & Nucl. Phys. A \textbf{258}, 341 (1976) & $^{25}$Al, $^{37}$K \\

    [Mo71] & C.E. Moss \emph{et al.} & Nucl.Phys. A \textbf{170}, 111 (1971) & $^{25}$Al, $^{37}$K \\

    [Oi97] & M. Oinonen \emph{et al.} & Phys.Rev. C \textbf{56}, 745 (1997) &  $^{71}$Kr \\

    [Sa93] & E.R.J. Saettler \emph{et al.} & Phys. Rev. C \textbf{48}, 3069 (1993) &  $^{19}$Ne \\

    [Se96] & D.R. Semon \emph{et al.} & Phys.Rev. C \textbf{53}, 96 (1996) &  $^{57}$Cu \\

    [Ta60] & W.L. Talbert, Jr. and M.G. Stewart & Phys. Rev. \textbf{119}, 272 (1960) & $^{23}$Mg, $^{25}$Al, $^{27}$Si, $^{31}$S, $^{39}$Ca \\

    [Ti87] & D.R. Tilleya \emph{et al.} & Nucl. Phys. A \textbf{474}, 1 (1987) &    $^{3}H$ \\

    [We02] & L. Weissman \emph{et al.} & Phys.Rev. C \textbf{65}, art. No. 044321 &  $^{61}$Ga \\

    [Wi69] & G.L. Wick \emph{et al.} & Nucl.Phys. A \textbf{138}, 209 (1969) &  $^{35}$Ar \\

    [Wi80] & H.S. Wilson \emph{et al.} & Phys.Rev. C \textbf{22}, 1696 (1980) & $^{21}$Na, $^{25}$Al, $^{29}$P, $^{31}$S, $^{33}$Cl, $^{35}$Ar, $^{41}$Sc \\

\hline \hline
\end{tabular}

\caption{ References to data used in the calculation of the
various branching ratios.\label{usedref_BR}}

\end{table}

\newpage
\begin{table}[!h]

\begin{tabular}{llp{8cm}l}
\hline \hline
 & & & \\
 Code & Authors & Reference & Nucleus \\

 \hline
  & & & \\

\multicolumn{4}{l}{\bf No error bar quoted}            \\

    [St59] & R.S. Storey, K.G. McNeill & Can. J. Phys. \textbf{37}, 1072 (1959) &  $^{23}$Mg \\

    [Va60] & S.S. Vasilev, L.Y. Shavtvalov & Zhur. Eksptl. I teoret. Fiz. \textbf{39}, 1221 (1960), Soviet Phys. JETP \textbf{12}, 851 (1961) &  $^{27}$Si \\

    [Ma76] & F.M. Mann \emph{et al.} & Nucl. Phys. A \textbf{258}, 341 (1976) &  $^{39}$Ca \\

    [En73] & P.M. Endt, C. van der Leun & Nucl. Phys. A \textbf{214}, 1 (1973) &  $^{43}$Ti \\

    [Ko73] & S. Kochan \emph{et al.} & Nucl. Phys. A \textbf{204}, 185 (1973) &  $^{53}$Co \\

    [Ho77] & P. Hornshoj \emph{et al.} & Nucl. Phys. A \textbf{288}, 429 (1977) &  $^{55}$Ni \\

    [Ay84] & J. \"{A}yst\"{o} \emph{et al.} & Phys. Lett. B \textbf{138}, 369-372 (1984) &  $^{55}$Ni \\

    [Ar81] & Y. Arai \emph{et al.} & Phys. Lett. B \textbf{104},186 (1981) &  $^{59}$Zn \\

    [Wi93] & J.A. Winger \emph{et al.} & Phys. Lett. B \textbf{299}, 214 (1993), Phys. Rev. C \textbf{48}, 3097 (1993) & $^{63}$Ge, $^{65}$As \\

    [Ba94] & P. Baumann \emph{et al.} & Phys. Rev. C \textbf{50}, 1180 (1994) &  $^{67}$Se \\

    [Bl95] & B. Blank \emph{et al.} & Phys. Lett. B \textbf{364}, 8 (1995) & $^{67}$Se, $^{71}$Kr \\

    [Ew81] & G.T. Ewan \emph{et al.} & Nucl. Phys. A \textbf{352}, 13 (1981) &  $^{71}$Kr \\

           &            &            &            \\

\multicolumn{4}{l}{\bf Error bar 10 times higher than most precise measurement}            \\

    [Ro55] & H. Roderick  \emph{et al.} & Phys. Rev. \textbf{97}, 97-101 (1955) &   $^{29}$P \\

    [Ki56] & O.C. Kistner \emph{et al.} & Phys. Rev. \textbf{104}, 154 (1956) &  $^{35}$Ar \\

    [Ad84] & E.G. Adelberger \emph{et al.} & Nucl. Phys. A \textbf{417}, 269 (1984) &  $^{39}$Ca \\

    [Oi99] & M. Oinonen \emph{et al.} & Eur. Phys. J. A \textbf{5}, 151 (1999) &  $^{61}$Ga \\

           &            &            &            \\

\multicolumn{4}{l}{\bf No branching ratio is given, only a (lower) limit}           \\

    [Ki58] & O.C. Kistner, B.M. Rustad & Phys. Rev. \textbf{112}, 1972 (1958) &  $^{39}$Ca \\

    [De71] & C. D\'{e}traz \emph{et al.} & Phys. Lett. B \textbf{34},128 (1971) &  $^{39}$Ca \\

    [Az77] & G. Azuelos \emph{et al.} & Phys. Rev. C \textbf{15}, 1847 (1977) &  $^{39}$Ca \\

    [Ho77] & P. Hornshoj \emph{et al.} & Nucl. Phys. A \textbf{288}, 429 (1977) & $^{47}$Cr, $^{51}$Fe \\

           &            &            &            \\

\multicolumn{4}{l}{\bf $\beta^+$ contamination from $^{21}$F according to [Al74]}          \\

    [Ta60] & W.L. Talbert, Jr. and M.G. Stewart & Phys. Rev. \textbf{119}, 272 (1960) &  $^{21}$Na \\

    [Ar64] & S.E. Arnell, E. Wernbom & Arkiv Fysik \textbf{25}, 389 (1964) &  $^{21}$Na \\

           &            &            &            \\

\multicolumn{4}{l}{\bf Only one important level in daughter was used, while there are more}           \\

    [Sh84] & T. Shinozuka \emph{et al.} & Phys. Rev. C \textbf{30}, 2111 (1984) &  $^{57}$Cu \\

\hline \hline

\end{tabular}

\caption{ References to data that were not used in the calculation
of the branching ratios, with the reason for their
rejection.\label{unusedref_BR} }

\end{table}

\end{document}